\documentclass[twocolumn]{aastex631}
\usepackage{mathtools, cuted}
\usepackage{lipsum, color}

\shorttitle{Finding universal relations in subhalo properties with artificial intelligence}
\shortauthors{H. Shao, F Villaescusa-Navarro et al.}

\graphicspath{{./}{figures/}}

\begin{document}

\title{Finding universal relations in subhalo properties with artificial intelligence}

\author{Helen Shao}
\altaffiliation{hshao@princeton.edu}
\affiliation{Department of Astrophysical Sciences, Princeton University, Peyton Hall, Princeton NJ 08544, USA}

\author{Francisco Villaescusa-Navarro}
\altaffiliation{villaescusa.francisco@gmail.com}
\affiliation{Department of Astrophysical Sciences, Princeton University, Peyton Hall, Princeton NJ 08544, USA}
\affiliation{Center for Computational Astrophysics, Flatiron Institute, 162 5th Avenue, New York, NY, 10010, USA}

\author{Shy Genel}
\affiliation{Center for Computational Astrophysics, Flatiron Institute, 162 5th Avenue, New York, NY, 10010, USA}
\affiliation{Columbia Astrophysics Laboratory, Columbia University, New York, NY, 10027, USA}

\author{David N. Spergel}
\affiliation{Center for Computational Astrophysics, Flatiron Institute, 162 5th Avenue, New York, NY, 10010, USA}
\affiliation{Department of Astrophysical Sciences, Princeton University, Peyton Hall, Princeton NJ 08544, USA}

\author{Daniel Angl\'es-Alc\'azar}
\affiliation{Department of Physics, University of Connecticut, 196 Auditorium Road, U-3046, Storrs, CT, 06269, USA}
\affiliation{Center for Computational Astrophysics, Flatiron Institute, 162 5th Avenue, New York, NY, 10010, USA}

\author{Lars Hernquist}
\affiliation{Center for Astrophysics | Harvard \& Smithsonian, 60 Garden St., Cambridge, MA 02138, USA}

\author{Romeel Dav\'e}
\affiliation{Institute for Astronomy, University of Edinburgh, Royal Observatory, Edinburgh EH9 3HJ, UK}
\affiliation{Department of Physics \& Astronomy, University of the Western Cape, Cape Town 7535,
South Africa}
\affiliation{South African Astronomical Observatories, Observatory, Cape Town 7925, South Africa}

\author{Desika Narayanan}
\affiliation{Department of Astronomy, University of Florida, Gainesville, FL, USA}
\affiliation{University of Florida Informatics Institute, 432 Newell Drive, CISE Bldg E251, Gainesville, FL, USA}

\author{Gabriella Contardo}
\affiliation{Center for Computational Astrophysics, Flatiron Institute, 162 5th Avenue, New York, NY, 10010, USA}

\author{Mark Vogelsberger}
\affiliation{Kavli Institute for Astrophysics and Space Research, Department of Physics, MIT, Cambridge, MA 02139, USA}

\begin{abstract}
We use a generic formalism designed to search for relations in high-dimensional spaces to determine if the total mass of a subhalo can be predicted from other internal properties such as velocity dispersion, radius, or star-formation rate. We train neural networks using data from the Cosmology and Astrophysics with MachinE Learning Simulations (CAMELS) project and show that the model can predict the total mass of a subhalo with high accuracy: more than 99\% of the subhalos have a predicted mass within 0.2 dex of their true value. The networks exhibit surprising extrapolation properties, being able to accurately predict the total mass of any type of subhalo containing any kind of galaxy at any redshift from simulations with different cosmologies, astrophysics models, subgrid physics, volumes, and resolutions, indicating that the network may have found a universal relation. We then use different methods to find equations that approximate the relation found by the networks and derive new analytic expressions that predict the total mass of a subhalo from its radius, velocity dispersion, and maximum circular velocity. We show that in some regimes, the analytic expressions are more accurate than the neural networks. We interpret the relation found by the neural network and approximated by the analytic equation as being connected to the virial theorem.
\end{abstract}


\section{Introduction}
\label{sec:intro}

Galaxies are fascinating objects that host dark matter, gas, stars, and black-holes. These elements interact in rich and complex manners whose details we do not yet fully understand \citep{SomervilleDave2015}. Improving our knowledge on the processes and the physics driving galaxy formation and evolution is not only important for galactic physics, but will also benefit other fields such as cosmology, where the uncertainty in galactic physics represents a major obstacle in extracting fundamental physics information from cosmic surveys \citep{CAMELS}.

Galaxies are characterized by many different properties, such as stellar mass, gas metallicity, neutral hydrogen mass, and luminosity in a given band. However, most of these properties are not independent and there are well-known correlations among them. In some cases, the correlations are tight enough to become actual relations such as the Tully-Fisher relation \citep{Tully_Fisher}, the Faber-Jackson relation \citep{faberjackson}, and the fundamental plane defined by velocity dispersion, effective radius, and effective surface brightness \citep{dressler, stellar_systems, galaxy_properties}. Some of these correlations and relations are induced by physical mechanisms, and therefore their existence reflects the underlying laws governing a particular process. In that regard, it is important to search for such relationships in high-dimensional spaces, like the one concerning galaxies, as those relations can help us learn about the underlying physics. Unfortunately, it is not easy to find relations in high-dimensional spaces. On the other hand, machine learning techniques can perform this task in a relatively straightforward way. 

In this work we outline a generic methodology that can be used to search for such relations in high-dimensional spaces and apply it to the case of the properties of subhalos and the galaxies within them. We identify a potentially new and universal relation between the total mass of a subhalo and other internal properties such as radius, velocity dispersion, and gas metallicity by training neural networks on subhalo properties from state-of-the-art hydrodynamic simulations. We then derive a new analytic expression that is able to parametrize the found relationship in an accurate way.

This paper is organized as follows. In Section \ref{sec:methods} we describe the data and the machine learning algorithm we use. We then present our findings in Section \ref{sec:results}. Finally, we draw the main conclusions of this work in Section \ref{sec:conclusions}.

\section{Methods}
\label{sec:methods}

In this work we search for a relation between the total mass of a subhalo and its other properties. In this section we first outline the generic methodology we employ to search for relationships between different variables in high dimensional spaces and how to identify analytic expressions that approximate them. We then describe in detail the ingredients needed to carry out this task: the data used, the neural network architecture and training procedure, and the method used to find the analytic equations.

\subsection{Methodology}
\label{subsec:methodology}

The generic idea behind this method is to find a relation between a given variable, $y$, and a set of other variables $\vec{x}=\{x_1, x_2,...,x_N\}$, i.e.
\begin{equation}
    y = f(\vec{x})
\end{equation}
In the first step, one can train a neural network to approximate the function $f$. By testing the accuracy of the network on the test set, one can quantify the accuracy reached by the model. If the accuracy is high, there may be an actual relationship between the variables. One can then use methods such as saliency maps to identify the most important variables that contribute to the relationship. By training neural networks using only the most important variables, one can check if the model extrapolates better or not than the model trained on all properties; in other words, whether the model may be extracting spurious information from some variables. Next, if the set of variables that contribute the most and achieve a good accuracy is small enough, one can use techniques such as symbolic regression to approximate the found relationship with analytic expressions. If the derived expression is not accurate enough, one can try with simpler models (e.g. power laws) guided by physical principles (if possible). For instance, the symbolic regression algorithm may find equations that, although not perfect, they may capture the main trend, and one can improve on them based on physical principles or by adding additional dependencies not captured by the model.

We apply the above scheme to internal subhalo properties from state-of-the-art hydrodynamic simulations, and derive a new relation between the total mass of a subhalo and its other internal properties and also the properties of the galaxy it hosts. We then provide an equation that approximates such a relation and attempt to understand the physics behind it.

\subsection{Data}

We made use of data from IllustrisTNG100 and IllustrisTNG300 simulations \citep{MarinacciF_17a,NaimanJ_17a,NelsonD_17a,IllustrisTNG_public,SpringelV_17a,Pillepich_2018} together with the CAMELS project \citep{CAMELS}, that we briefly describe now.

\begin{itemize}
\item \textbf{IllustrisTNG100}. A state-of-the-art magneto-hydrodynamic simulation run with the moving mesh AREPO code \citep{Arepo_public}. The simulation samples a periodic comoving volume of $(75~h^{-1}{\rm Mpc})^3$ using 1,820$^3$ dark matter particles and 1,820$^3$ fluid elements down to $z=0$. The details of the subgrid galaxy formation model used in this simulation can be found in \citet{PillepichA_16a, WeinbergerR_16a}, including supernova-driven galactic winds and Active Galactic Nuclei (AGN) feedback. Out of the three simulations run at different resolutions, we use the one with the highest resolution: IllustrisTNG100-1. This simulation has the highest mass and spatial resolution of the ones considered in this work. We use this simulation to quantify how our networks and analytic expressions behave for subhalos from a simulation with higher resolution and larger volume than the ones they have been trained on.

\item \textbf{IllustrisTNG300}. This simulation is identical to IllustrisTNG100 with the only difference being its volume, $(205~h^{-1}{\rm Mpc})^3$ and its resolution, with 2,500$^3$ dark matter particles plus 2,500$^3$ fluid elements. Out of the three simulations run at different resolutions, we use the one with the highest resolution: IllustrisTNG300-1. We use this simulation to quantify how our networks and analytic expressions behave for subhalos from a simulation with a much larger volume and slightly different resolution than the ones they have been trained on.

\item \textbf{CAMELS-IllustrisTNG}. A suite of 1,000 magneto-hydrodynamic simulations run with the AREPO code and using the same subgrid model as the IllustrisTNG simulations above. Each simulation follows the evolution of $256^3$ dark matter particles plus $256^3$ fluid elements in a periodic comoving volume of $(25~h^{-1}{\rm Mpc})^3$ down to $z=0$. Each of the 1,000 simulations has a different cosmology (varying $\Omega_{\rm m}$ and $\sigma_8$) but also astrophysics by varying the value of four astrophysical parameters controlling the efficiency of supernova and AGN feedback. These simulations are part of the CAMELS project \citep{CAMELS}. We use the subhalos from these simulations to train the neural networks.

\item \textbf{CAMELS-SIMBA}. A suite of 1,000 hydrodynamic simulations run with the GIZMO code \citep{Hopkins2015_Gizmo} employing the same subgrid model as the SIMBA simulation \citep{SIMBA}. 
These simulations are comparable to CAMELS-IllustrisTNG in the sense that each simulation follows the evolution of $256^3$ dark matter particles plus $256^3$ fluid elements in a periodic comoving volume of $(25~h^{-1}{\rm Mpc})^3$ down to $z=0$. Each of the 1,000 simulations has a different cosmology (varying $\Omega_{\rm m}$ and $\sigma_8$) but also astrophysics by varying the value of four astrophysical parameters controlling the efficiency of supernova and AGN feedback. We emphasize that the subgrid physics of SIMBA is very different to that of IllustrisTNG, including parameterized galactic winds based on higher resolution FIRE simulations \citep{Muratov2015,Angles-Alcazar2017_BaryonCycle} and gravitational torque-driven black hole growth coupled to kinetic outflows \citep{Angles-Alcazar2017_BHfeedback}. These simulations are part of the CAMELS project \citep{CAMELS}. We use the subhalos of these simulations to quantify the accuracy of our networks and analytic expressions when varying the subgrid physics and method used to solve the hydrodynamic equations.

\item \textbf{CAMELS-Nbody}. Each hydrodynamic simulation belonging to CAMELS-IllustrisTNG and CAMELS-SIMBA has an associated gravity-only N-body simulation with the same cosmological model and the same initial random phase as its hydrodynamic counterpart. These simulations thus follow the evolution of $256^3$ dark matter particles in a periodic comoving volume of $(25~h^{-1}{\rm Mpc})^3$ down to $z=0$. These simulations only account for the gravitational forces and therefore do not model galaxy formation physics. We use these simulations to check if the findings from the above hydrodynamic simulations also apply to subhalos from N-body simulations.

\end{itemize}
Table \ref{tab:simulations} summarizes the main characteristics of the different simulations used in this work. Each simulation has a series of halo and subhalo catalogues associated to it that were generated by running \textsc{SUBFIND} \citep{Subfind} on the corresponding snapshots at different redshifts. \textsc{SUBFIND} works primarily by identifying local peaks in the three-dimensional density field and separating them by identifying a saddle point between them. In the second step, the overdensities and their surroundings are checked for gravitational self-boundness: those that are self-bound are registered as subhalos, and those that are not are attached to their neighboring overdensities, namely those they share saddle points with. Subfind operates on all particle types in the simulations, dark matter and baryonic alike. In this work we focus our attention on subhalos that contain \textit{galaxies}, i.e. \textsc{SUBFIND} subhalos that contain more than 20 star particles. For each subhalo we consider twelve different quantities including both properties of the subhalo itself and of its central galaxy: 
\begin{itemize}
\item \textbf{Total mass}, $M_{\rm tot}$. This quantity represents the total mass of the subhalo, i.e. the sum of the mass in gas, dark matter, stars, and black-holes. 
\item \textbf{Black-hole mass}, $M_{\rm BH}$. This quantity is the black-hole mass of the subhalo. 
\item \textbf{Stellar mass}, $M_*$. This quantity is the stellar mass of the subhalo. 
\item \textbf{Gas mass}, $M_{\rm g}$. This quantity represents the gas mass of the subhalo. 
\item \textbf{Spin}, $J$. This quantity is the modulus of the subhalo 3D spin vector. 
\item \textbf{Velocity}, $V$. The quantity is the modulus of the peculiar velocity vector of the subhalo.
\item \textbf{Gas metallicity}, $Z_g$. This quantity represents the average metallicity of all gas particles within twice the radius containing half of the subhalo stellar mass. 
\item \textbf{Stars metallicity}, $Z_*$. This quantity is the average metallicity of all star particles within twice the radius containing half of the subhalo stellar mass. 
\item \textbf{Radius}, $R$. This quantity is the comoving radius containing half of the subhalo total mass. 
\item \textbf{Star formation rate}, ${\rm SFR}$. This quantity is the total star formation rate of all gas particles in the subhalo. 
\item \textbf{Velocity dispersion}, $\sigma$. This quantity represents the 1D velocity dispersion of all particles in the subhalo. 
\item \textbf{Maximum velocity curve}, $V_{\rm max}$. This quantity is the maximum of the spherically-averaged rotation curve, defined as $\sqrt{GM_{\rm tot}(<r)/r}$. 
\end{itemize}
We note that some of the above properties can be associated with the galaxy inside the subhalo, like the black-hole mass, stellar mass, and star-formation rate. Others, like the gas mass can be associated with the galaxy but also its circum-galactic medium. We emphasize that although we generically call all these objects subhalos, some of them are may be very close to actual halos. For instance, the most massive subhalo of a halo tends to represent the main body of the dark matter halo without including other subhalos and their galaxies. Thus, while we do not distinguish between central and satellites in our nomenclature (but see Appendix \ref{sec:centrals_satellites}), it is important to keep this in mind when associating our subhalos to the physical systems behind them.

\begin{table*}[ht!]
\begin{center}
\begin{tabular}{|l|c|c|p{6cm}|}
\hline
\textbf{Simulation Name} & \textbf{Number of Subhalos} & \textbf{$M_{\rm tot}~[h^{-1}M_\odot]$} & \textbf{Description}\\
\hline \hline 
IllustristTNG100    & 88,507 & $[8.30\times10^6 - 2.75\times10^{14}]$ & A hydrodynamic simulation run with the AREPO code. The simulation has a volume of $(75~h^{-1}{\rm Mpc})^3$.  \\
\hline
IllustristTNG300    & 515,600 & $[6.89\times10^7 - 1.28\times10^{15}]$ & A hydrodynamic simulation run with the AREPO code. The simulation has a volume of $(205~h^{-1}{\rm Mpc})^3$. \\
\hline
CAMELS-IllustrisTNG & 720,548 &  $[1.18\times10^8 -  3.89\times10^{14}]$  & A set of 1,000 simulations run with the AREPO code employing the same subgrid physics as the IllustrisTNG simulations. Each simulation has a different value of the cosmological and astrophysical parameters. Each simulation has a volume of $(25~h^{-1}{\rm Mpc})^3$.  \\
\hline
CAMELS-SIMBA     & 1,182,265 & $[1.55\times10^{8} - 5.36\times10^{14}]$ & A set of 1,000 simulations rut with the GIZMO code employing the same subgrid physics as the SIMBA simulations. Each simulation has a different value of the cosmological and astrophysical parameters.  Each simulation has a volume of $(25~h^{-1}{\rm Mpc})^3$.  \\
\hline
CAMELS-Nbody     & 16,965,513 & $[5.44\times10^{8} - 3.93\times10^{14}]$ & A set of 2,000 N-body simulations run with the Gadget-III code. Each simulation has a different value of the cosmological parameters, and they represent the gravity-only counterpart of the CAMELS-IllustrisTNG and CAMELS-SIMBA simulations. Each simulation has a volume of $(25~h^{-1}{\rm Mpc})^3$.  \\
\hline
\end{tabular}
\end{center}
\caption{This table lists the names of the simulations we use, the number of subhalos they contain, and their maximum and minimum subhalo masses. For the hydrodynamic simulations, we only consider subhalos that contain more than 20 star particles. For the N-body simulations, we only consider subhalos that contain more than 20 dark matter particles. \label{tab:simulations}}
\end{table*}

\subsection{Neural networks}
\label{subsec:NN}

The main goal of this paper is to find a relation between the total mass of a subhalo, $M_{\rm tot}$, and its other properties, i.e.
\begin{equation}
M_{\rm tot}=f(\vec{\theta})
\end{equation}
where $\vec{\theta}$ is the vector with the other subhalo properties; $\vec{\theta}$ can be all other subhalo properties or a subset of it. Neural networks can be used to approximate the function $f$. 
When training a model on a given dataset, we first split its data into training (80\%), validation (10\%), and testing (10\%). Some quantities exhibit a very large dynamical range. This can be a problem when training the networks. Thus, we first redefine these variables such as: 
\begin{itemize}
\item $M_{\rm tot}\longrightarrow\log_{10}(1+M_{\rm tot})$
\item $M_*\longrightarrow\log_{10}(1+M_*)$
\item $M_{\rm BH}\longrightarrow\log_{10}(1+M_{\rm BH})$
\item $M_g\longrightarrow\log_{10}(1+M_g)$
\end{itemize}
Before taking the logarithm we have added one to the variables to avoid problems when the value of those variables is equal to 0. Note that we are implicitly assuming that each mass term is divided by $h^{-1}M_\odot$. We then standardize all variables using:
\begin{equation}
    \tilde{x}=\frac{x - {\mu}}{{\delta}}
\end{equation}
where ${\mu}$ is the mean and ${\delta}$ is the standard deviation of each variable, $x$.

The architecture of our model consists of a set of fully connected layers whose input is the vector with the considered normalized variables and the output is a single number with the predicted value of the normalized subhalo total mass, $\tilde{M}_{\rm tot}^{\rm }$. Between each fully connected layer we include a LeakyReLU non-linear activation function with a negative slope value of 0.2.

The loss function we optimize via gradient descent is the standard mean squared error:
\begin{equation}
    \mathcal{L}=\frac{1}{N}\sum_{i=1}^N \left(\tilde{M}_{\rm tot}^{\rm true}-\tilde{M}_{\rm tot}^{\rm pred}\right)^2
\end{equation}
We use the AdamW optimizer \citep{AdamW} with beta values equal to 0.9 and 0.999. We train the network using a batch size of 256 for 500 epochs. We use \textsc{Pytorch}\footnote{\url{https://pytorch.org}} to train, validate, and test the networks.

The hyper-parameters of our model are 1) the number of fully connected layers, 2) the number of neurons per layer, 3) the value of the learning rate, and 4) the weight decay. We use the \textsc{optuna} code \citep{Optuna} to perform Bayesian optimization and find the best-value of these hyper-parameters for each case we consider (e.g. when considering all subhalo properties or when considering only a subset of them). For each case we run 100 trials, where each trial consists of training the model using selected values of the hyper-parameters. We perform the optimization of the hyper-parameters requiring to achieve the lowest validation loss possible. The model we choose in the end is the one with the lowest validation loss found by \textsc{optuna}.

\subsection{Symbolic regression}
\label{subsec:SR}

While neural networks can approximate very complex relations hidden in the data, their interpretation may be challenging. In general, it is desirable to obtain an expression that characterizes, or approximates, a given relation, because the physics behind it is much easier to understand and interpret in that form. For this purpose we made use of techniques designed to approximate functions with analytic expressions.

\begin{figure*}[ht!]
    \centering
    \includegraphics[width=1\textwidth]{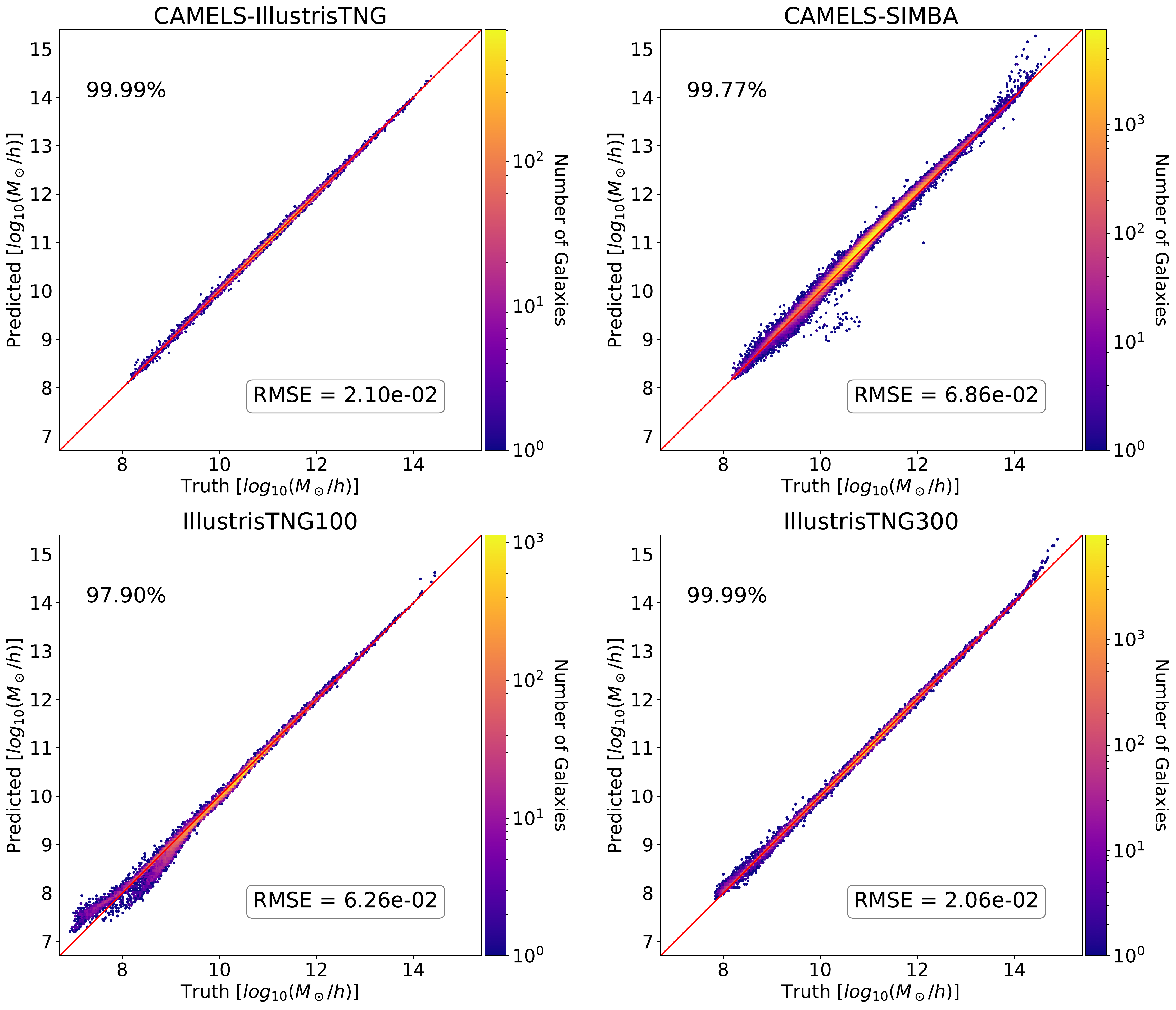}
    \caption{We train a neural network to predict the total mass of a subhalo from other 11 properties of the subhalo and the galaxy that it hosts. The training is carried out using subhalos from the CAMELS-IllustrisTNG simulations at redshift $z=0$.    The different panels show the predicted mass versus the true value for subhalos from the CAMELS-IllustrisTNG (top-left), CAMELS-SIMBA (top-right), IllustrisTNG100 (bottom-left), and IllustrisTNG300 (bottom-right) simulations. The plots are 2D histograms with the color bar indicating the number of galaxies in each bin of predicted and true mass. The root mean square error is quoted on the bottom right of each plot and the percentage of the predictions that lie within 0.2 dex of the actual value is quoted on the top left corners. The trained model exhibits surprising extrapolation properties, indicating that a universal relation may have been found.}
    \label{fig:nn_all}
\end{figure*}

We made use of two different methods: 
\begin{itemize}
\item \textbf{Genetic programming.} The idea behind this method is to start with a series of operators (e.g. $+,-,\times,/,\exp$) and create combinations with the operators and the variables in a so-called generation. Those expressions are then evaluated and the most accurate ones \textit{survive} to the next generation, where \textit{mutations} and \textit{crossovers} can take place to explore different equations and find an optimal one. We made use of the \textsc{pysr}\footnote{https://pysr.readthedocs.io} and \textsc{Eureqa}\footnote{https://www.creativemachineslab.com/eureqa.html} packages for this.

\item \textbf{Template fitting.} Differently from genetic programming, here we start with a functional form that has some free parameters and perform a non-linear least squares fit to the data to obtain the best value of the parameters. This is done using the curve fit function from \textsc{scipy-optimize}\footnote{https://docs.scipy.org/doc/scipy/reference/optimize.html}. The idea is to use physically motivated expressions or to improve the results obtained from genetic programming. 
\end{itemize}
When searching for analytic expressions we typically combine both methods. For instance, we use genetic programming to obtain the main trend and then we improve on that using template fitting.

\section{Results} 
\label{sec:results}

In this section, we present the results of our analysis with the neural network and symbolic regression models.

\subsection{Neural Networks} 
\label{subsec:nn}

We start by training a neural network to predict the total subhalo mass from the other 11 properties using the subhalos of the CAMELS-IllustrisTNG simulations. In the top-left panel of Fig. \ref{fig:nn_all} we show the predicted total subhalo mass versus its true value from the subhalos of the CAMELS-IllustrisTNG test set (that the network has not seen before). As can be seen, the network is able to predict the value of the subhalo total mass very accurately: the RMSE is $2.10\times10^{-2}$ and 99.99\% of the subhalos have a predicted total mass that is within 0.2 dex of its true value. 

One may wonder if the tight relation found by the network arises from a strong correlation with a single variable. For instance, the network may have found a trivial relation between the mass and the radius of a subhalo. To check this, we make scatter plots between $M_{\rm tot}$ and $R$, $M_{\rm tot}$ and $\sigma$, and $M_{\rm tot}$ and $V_{\rm max}$ - as these three are the most important ones used by the network (see below) - and fit a generic function that goes through the mean of the relation between the variables. Given that function, one can predict the total subhalo mass given the value of the considered variable. The RMSE that these functions return are $3.48 \times 10^{-1}$ (for $R$), $3.79 \times 10^{-1}$ (for ${\sigma}$), and $4.48\times 10^{-1}$ (for $V_{\rm max}$). These RMSE are much higher than the one obtained by the network, indicating that the relation cannot be attributed to a strong correlation with a single variable.

We then test the model on the subhalos from the CAMELS-SIMBA, IllustrisTNG100, and IllustrisTNG300 simulations and show the results in the other panels of Fig. \ref{fig:nn_all}. We find that even if these subhalos come from simulations with different subhalo mass ranges, resolutions, hydrodynamic solvers and subgrid models than the ones used for training, the model can still accurately predict the total mass of those subhalos. The RMSE for the IllustrisTNG300 subhalos is similar to the one from the CAMELS-IllustrisTNG, while for CAMELS-SIMBA and IllustrisTNG100 the RMSE is a factor of $\simeq3$ larger. We note that out of these three, it is expected that the lowest RMSE will be obtained by the subhalos from the IllustrisTNG300 simulation, as those subhalos are similar to those from the IllustrisTNG-CAMELS simulations both in terms of resolution and simulation code. The fraction of subhalos whose predicted total mass is within 0.2 dex of the truth is very high in all cases: 99.77\% (CAMELS-SIMBA), 97.90\% (IllustrisTNG100), and 99.99\% (IllustrisTNG300). 

We note however a few interesting points. First, the scatter in the predicted total mass is significantly larger for CAMELS-SIMBA subhalos than for subhalos of the other simulations for almost all masses. Second, the network is not able to accurately predict the total mass of the most massive subhalos. Third, the network predicts the total subhalo mass with large scatter for very small subhalos; e.g. those with masses below $10^8~h^{-1}M_\odot$ from the IllustrisTNG100 simulation. All these cases can be seen as the network not extrapolating properly. In the case of CAMELS-SIMBA, their subhalo's galaxies are generated with a code that solves the hydrodynamic equations in a different way and utilizes a distinct subgrid model. In the case of IllustrisTNG100 and IllustrisTNG300, their higher resolution and larger volume allow these simulations to contain subhalos with lower and higher total mass than those in the CAMELS-IllustrisTNG simulations used for training (see Table \ref{tab:simulations}). Nevertheless, the model overall exhibits superb accuracy and appears to utilize a robust relation between different internal subhalo properties to predict the total mass.

We also examine how well the neural network is able to predict the total mass for central versus satellite subhalos, as these two types of systems exhibit different physical characteristics. We find that for all four simulations, the neural network predictions for the total mass of central subhalos demonstrate slightly higher accuracy than those for the satellite subalos. The RMSE for the centrals (satellites) are:  for CAMELS-IllustrisTNG, $1.94\times10^{-2}$ ($2.26\times10^{-2}$); for CAMELS-SIMBA, $6.69\times10^{-2}$ ($7.11\times10^{-2}$), for IllustrisTNG100, $3.97\times10^{-2}$ ($7.91\times10^{-2}$), and for IllustrisTNG300 $1.90\times10^{-2}$ ($2.25\times10^{-2}$). In general, the neural network effectively captures the trend for the total mass for both types of subhalos. We provide more details in Appendix \ref{sec:centrals_satellites}.

Our model not only works for subhalos from simulations with different volumes, resolutions, and subgrid models to the ones used for training, but also at redshifts other than the ones used for training. We illustrate this in Appendix \ref{sec:redshift}, where we display in Fig. \ref{fig:redshifts} the results of testing the network on redshifts other than $z=0$ (the one used for training). Our model is able to extrapolate very well, and the RMSE and fraction of subhalos within 0.2 dex is similar to those at $z=0$ and higher redshifts for subhalos from all simulation types. The largest differences we find are for subhalos of the CAMELS-SIMBA simulations. 

One key aspect needed for the network to work at higher redshifts was to use the proper radius, $R_p$, instead of the comoving radius, $R = R_p/a$, where $a=1/(1+z)$ is the scale factor. Without this change of variable, the network was unable to predict the correct total mass of the subhalos. This indicates that the relation learned by the network depends on the physical radius rather than the comoving radius. We provide further details on this issue in Appendix \ref{sec:variables}. 

\begin{figure}
    \includegraphics[width=0.48\textwidth]{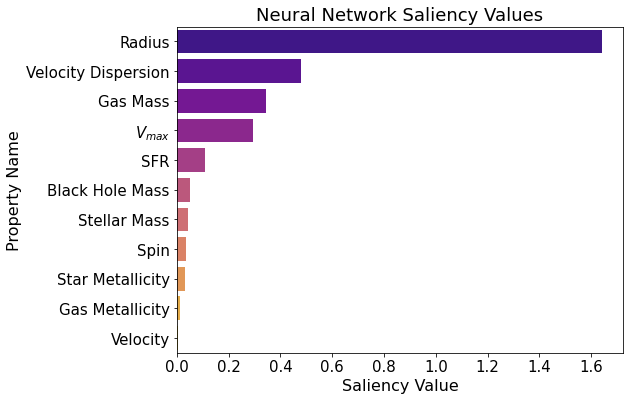}
    \caption{The saliency values of the input subhalo properties. Larger saliency values indicate that the network output is more affected by changes in the value of that variable and it can be seen as a proxy for feature importance.
    As shown, the five most important properties are $R$,  ${\sigma}$, $V_{\rm max}$, $M_g$, and ${\rm SFR}$.}
    \label{fig:saliency}
\end{figure}

Next, we examine which subhalo properties are the most important for the neural network's predictions by computing the saliency value of each input variable. This is shown in Figure \ref{fig:saliency}; we find that the top five most important properties, in descending order, are: $R$, ${\sigma}$, $V_{\rm max}$, $M_g$, and ${\rm SFR}$. We note that this exercise is done for subhalos with different masses, and the saliency values represents the average importance over all different subhalo masses; subhalos with different masses may exhibit different saliency values for the different variables.

To explore the effect of these variables on the accuracy of the neural network, we train four separate models to predict the total subhalo mass using different combinations of the five most important properties and compare the results with that of the original model that was trained on all eleven properties. We list the RMSE of the predictions from each of the models when tested on subhalos from the four different simulations in Table \ref{tab:nn_table}. We find that for subhalos of the CAMELS-IllustrisTNG, training on more variables helps in achieving higher accuracy. This is expected since adding more variables should only increase the information content, not decrease it. We observe the same behavior when the network is tested on subhalos from the IllustrisTNG300 simulation.

On the other hand, the model trained on the five subhalo properties achieves a higher accuracy than the model trained on all properties when tested using subhalos of the CAMELS-SIMBA and IllustrisTNG100 simulations. This may be due to the fact that the network, when trained on all properties, may be extracting information from properties that are unique to one particular simulation. In other words, the network may be learning artificial correlations that are only present in simulations with a given physics implementation or resolution. The model trained on 5 properties seems to be learning more robust relationships that are able to better extrapolate when applied to subhalos from other simulations.

\begin{deluxetable*}{|c|c|c|c|c|c|}
\tablewidth{0pt}
\tablecaption{We train neural networks to predict the total subhalo mass from different combinations of input properties using the subhalos from the CAMELS-IllustrisTNG simulations. Each column indicates which properties were used for training and each row shows the root mean square errors of the predicted total mass when the model was tested on the respective simulation sets. As seen, using more variables yields more accurate results for subhalos of the CAMELS-IllustrisTNG simulations, but in some cases it makes the extrapolation worse for subhalos of other simulations.
\label{tab:nn_table}}
\tablehead{
&All Properties& ${\sigma}$, $V_{\rm max}$, $
R$, $M_g$, ${\rm SFR}$ & ${\sigma}$, $V_{\rm max}$, $R$, $M_g$ & ${\sigma}$, $V_{\rm max}$, $R$ & ${\sigma}$, $R$}
\startdata
CAMELS-IllustrisTNG & $2.1\times10^{-2}$ & $2.3\times10^{-2}$ & $2.6\times10^{-2}$ & $3.0\times10^{-2}$ & $5.3\times10^{-2}$\\
\hline
CAMELS-SIMBA & $6.9\times10^{-2}$ & $5.2\times10^{-2}$ & $5.6\times10^{-2}$ & $6.6\times10^{-2}$ & $1.0\times10^{-1}$ \\
\hline
IllustrisTNG100 & $6.2\times10^{-2}$ & $5.3\times10^{-2}$ & $6.8\times10^{-2}$ & $9.0\times10^{-2}$ & $1.3\times10^{-1}$ \\
\hline
IllustrisTNG300 & $2.1\times10^{-2}$ & $2.2\times10^{-2}$ & $2.5\times10^{-2}$ & $3.0\times10^{-2}$ & $5.6\times10^{-2}$ \\
\hline
\enddata
\end{deluxetable*}

We also investigate if the relation found by the network is unique to subhalos of hydrodynamic simulations or whether it may be a more generic property that applies to subhalos of gravity-only N-body simulations. We carry out this task by testing the network trained on the $R$, $\sigma$, and $V_{\rm max}$ properties of the subhalos of the CAMELS-IllustrisTNG simulations on subhalos from N-body simulations of CAMELS. We show the results in Appendix \ref{sec:N-Body} (Fig. \ref{fig:N-Body}). We find that our model is also able to accurately predict the total mass of these subhalos that originate from simulations with different cosmological models. We provide further details of this test in Appendix \ref{sec:N-Body}.

Finally, we study if our conclusions change if we train on subhalos from simulations other than those of the CAMELS-IllustrisTNG simulations. For this, we train a network using all eleven properties from subhalos of the CAMELS-SIMBA simulations, and test the model on CAMELS-IllustrisTNG, IllustrisTNG100, and IllustrisTNG300. We find that that model is able to capture a similar trend for all simulations but predictions are less accurate than those from the model trained on CAMELS-IllustrisTNG subhalos. Similarly, we also train a neural network using all eleven properties of subhalos from the IllustrisTNG300 simulation and test the model on subhalos from other simulations. In this case we also find that the predictions have an order of magnitude higher RMSE than the predictions from the neural network trained on CAMELS-IllustrisTNG. We provide further details on these tests in Appendix \ref{sec:other_sims}. 

One reason that can explain these results is that the subhalos from the CAMELS-IllustrisTNG simulations exhibit a very large variety (given that they come from simulations with different cosmological and astrophysical parameters) that help the network to find more robust relations. When the network is trained on subhalos from a single simulation with fixed cosmology and astrophysics (e.g. IllustrisTNG300), the model may be learning artificial relations (such as numerical artifacts) that may not extrapolate very well. On the other hand, the subhalos from the IllustrisTNG100, IllustrisTNG300, and CAMELS-IllustrisTNG may share many similarities, since they are generated with the same code. Thus, the network trained on CAMELS-SIMBA subhalos may not extrapolate as well to these three simulations, in the same way that the model trained on CAMELS-IllustrisTNG does not extrapolate to subhalos from the CAMELS-SIMBA simulations as well as for the other IllustrisTNG-based simulations. Overall, although the model trained on CAMELS-IllustrisTNG seems to perform the best when testing on subhalos from other simulations, the models trained on CAMELS-SIMBA and IllustrisTNG300 are also very accurate and possess powerful extrapolation properties. We hence conclude that our results are not largely affected by the type of simulations used for training.

\subsection{Analytic expressions} \label{subsec:symbolic}

\begin{deluxetable*}{|c|c|c|c|c|}[t]
\tablewidth{0pt}
\tablecaption{The columns of this table list the root mean square errors of the various models when tested on the four simulations: CAMELS-IllustrisTNG, CAMELS-SIMBA, IllustrisTNG100, and IllustrisTNG300. Each row indicates the models defined in Equations \ref{Eq:simple_equation}, \ref{Eq:simple_equation2}, \ref{Eq:formula}, and \ref{Eq:eureqa}. \label{tab:equations_table}}
\tablehead{
Model & CAMELS-IllustrisTNG & CAMELS-SIMBA & IllustrisTNG100 & IllustrisTNG300}
\startdata
Equation \ref{Eq:simple_equation} & $7.60\times10^{-2}$ & $3.05\times10^{-1}$ & $1.24\times10^{-1}$ & $9.32\times10^{-2}$ \\
\hline
Equation \ref{Eq:simple_equation2} & $5.58\times10^{-2}$ & $2.02\times10^{-1}$ & $5.48\times10^{-2}$ & $4.94\times10^{-2}$ \\
\hline
Equation \ref{Eq:formula} & $3.96\times10^{-2}$ & $1.35\times10^{-1}$ & $3.08\times10^{-2}$ & $3.27\times10^{-2}$ \\
\hline
Equation \ref{Eq:eureqa} & $3.32\times10^{-2}$ & $1.07\times10^{-1}$ & $3.08\times10^{-2}$ & $2.74\times10^{-2}$ \\
\hline
\enddata
\end{deluxetable*}

\begin{figure*}[!ht]
    \centering
    \includegraphics[width=1\textwidth]{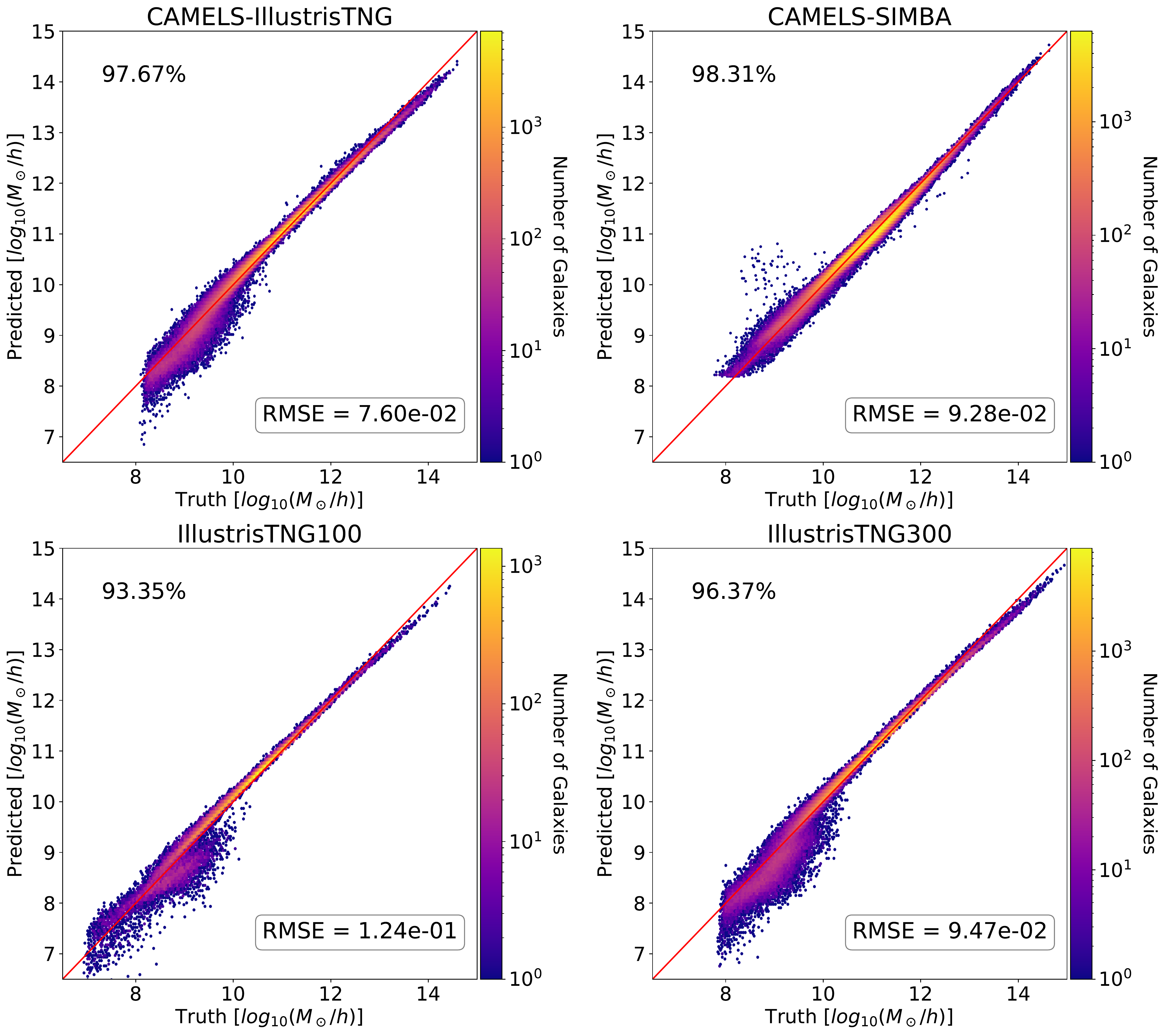}
    \caption{This figure shows the accuracy of the equation $M_{\rm tot}=A\sigma^\alpha R^\beta$ where $A = 10^{6.6}$, $\alpha = 1.9$, and $\beta = 0.9$, when used to predict the total mass of subhalos from the different simulations. Although this simple equation is able to capture the main trend of the relation, it can be seen that the predictions exhibit a large scatter in the low mass end and a significant bias in the high mass end for all simulations. The root mean square error is quoted in the bottom right and the percentage of predictions that lie within 0.2 dex of the actual value is quoted on the top left.}
    \label{fig:pysr_all}
\end{figure*}

The above results indicate that the function $M_{\rm tot}=f(\vec{\theta})$ found by the neural network may be a universal one, as it holds accurately when tested on subhalos from simulations with different cosmologies, astrophysical models, subgrid models and hydrodynamic solvers, different volumes, different resolutions, and at different redshifts than those used when training the model. If that is the case, it would be very interesting to obtain an analytic expression that reproduces, or approximates, $f$ in order to improve our knowledge on the physics behind such an relation.

In order to search for such an equation we first use genetic programming on a subset of inputs ($\vec{\theta}$) and outputs ($M_{\rm tot}$) from subhalos from the CAMELS-IllustrisTNG simulations that cover the whole mass range. Among the different equations found, the one that stands out for its simplicity and accuracy is
\begin{equation}
    M_{\rm tot}=A\sigma^\alpha R^\beta
    \label{Eq:simple_equation}
\end{equation}
where $A = 10^{6.6}$, $\alpha = 1.9$, and $\beta = 0.9$. We note that $A$ has units of $h^{-1}M_\odot$, while $\sigma$ and $R$ are assumed to have units of km/s and $h^{-1}{\rm kpc}$, respectively. This equation is able to capture the main trend of the function $f$ but it also exhibits significant biases and large scatter on the high and low mass ends, respectively (see Fig. \ref{fig:pysr_all}). The RMSE values of this equation are: $7.60\times10^{-2}$ (CAMELS-IllustrisTNG), $9.28\times10^{-2}$ (CAMELS-SIMBA), $1.24\times10^{-1}$ (IllustrisTNG100), and $9.32\times10^{-2}$ (IllustrisTNG300). Evidently, this model does not exhibit an accuracy comparable to that of the neural network discussed in section \ref{subsec:nn}. 

In order to improve the accuracy of the above expression we perform template fitting on the CAMELS-IllustrisTNG simulations, considering a relation of the form
\begin{equation}
M_{\rm tot}=A\sigma^\alpha R^\beta V_{\rm max}^\gamma
\label{Eq:simple_equation2}
\end{equation}
where $A$, $\alpha$, $\beta$, and $\gamma$ are the free parameters we fit to the data. We find several interesting things. First, the above equation is more accurate than Eq. \ref{Eq:simple_equation}, but its overall performance is still not satisfactory; the root mean square errors for this model tested on all simulations are listed in Table \ref{tab:equations_table}. This is mainly because we find that the free coefficients exhibit some dependence with the total mass of the subhalo. We fix this by using a slightly more complicated equation where the exponents have some implicit dependence on the parameters. Second, the dependence on $V_{\rm max}$ is crucial to accurately predict $M_{\rm tot}$ for small subhalos ($M_{\rm tot}<10^{10}~h^{-1}M_\odot$). We note that we find this dependence to be negligible for subhalos above $\simeq10^{12}~h^{-1}M_\odot$. We believe that this is expected since massive galaxies tend to have their maximum circular velocity at small radii deep within the stellar body of the galaxy (e.g. ~in the bulge), and hence it is expected to be related more to the detailed dynamics at small radii than to the total mass of the subhalo.

Thus, we choose to improve Eq. \ref{Eq:simple_equation2} by using a template of the form
\begin{equation}
M_{\rm tot}=A\sigma^{(\alpha_0+\alpha_1\log\sigma)} R^{(\beta_0+\beta_1\log R)} V_{\rm max}^{(\gamma_0+\gamma_1\log V_{\rm max})}
\end{equation}
where $\log(x)=\log_{10}(x)$. Instead of fitting the above equation over the whole mass range, we fit it over three different mass ranges: 1) $M_{\rm tot}< 10^{10}~h^{-1}M_\odot$, 2) $10^{10}~h^{-1}M_\odot < M_{\rm tot}<10^{12}~h^{-1}M_\odot$, 3) $M_{\rm tot}>10^{12}~h^{-1}M_\odot$. The equations we derive are:
\begin{widetext}
\begin{equation}
    M_{\rm tot}= 
\begin{cases}
    10^{5.47}R^{(0.96 - 0.06\log(R/R_0))}\sigma^{(0.19 - 0.10\log(\sigma/\sigma_0))}V_{\rm max}^{1.94}, & \text{if } M_{\rm tot}\leq 10^{10}~h^{-1}M_\odot\\
    
    10^{5.32}R^{(0.87 + 0.06\log(R/R_0))}\sigma^{
    0.72 + 0.40\log(\sigma/\sigma_0)}V_{\rm max}^{(1.58 - 0.54\log(V_{\rm max}/V_{\rm max 0}))}, & \text{if } 10^{10}~h^{-1}M_\odot<M_{\rm tot}<10^{12}~h^{-1}M_\odot\\  
    
    10^{7.37}R^{0.96}\sigma^{(1.46 + 0.24\log(\sigma/\sigma_0))},              & \text{if } M_{\rm tot}\geq 10^{12}~h^{-1}M_\odot\\
\end{cases}
\label{Eq:formula}
\end{equation}
\end{widetext}

In the above equations, for $M_{\rm tot}< 10^{10}~h^{-1}M_\odot$, $R_0 = 2.41$ $h^{-1}{\rm kpc}$ and $\sigma_0 = 25.6$ km/s. For $10^{10}~h^{-1}M_\odot < M_{\rm tot}<10^{12}~h^{-1}M_\odot$, $R_0 = 30.84$ $h^{-1}{\rm kpc}$, $\sigma_0 = 63.25$ km/s, and $V_{\rm max 0} = 123.33$ km/s. For $M_{\rm tot}>10^{12}~h^{-1}M_\odot$, $\sigma_0 = 148.96$ km/s. We also note that units of $R$, $\sigma$, and $V_{\rm max}$ are $h^{-1}{\rm kpc}$, km/s, and km/s, respectively.

\begin{figure*}[ht!]
    \centering
    \includegraphics[width=1\textwidth]{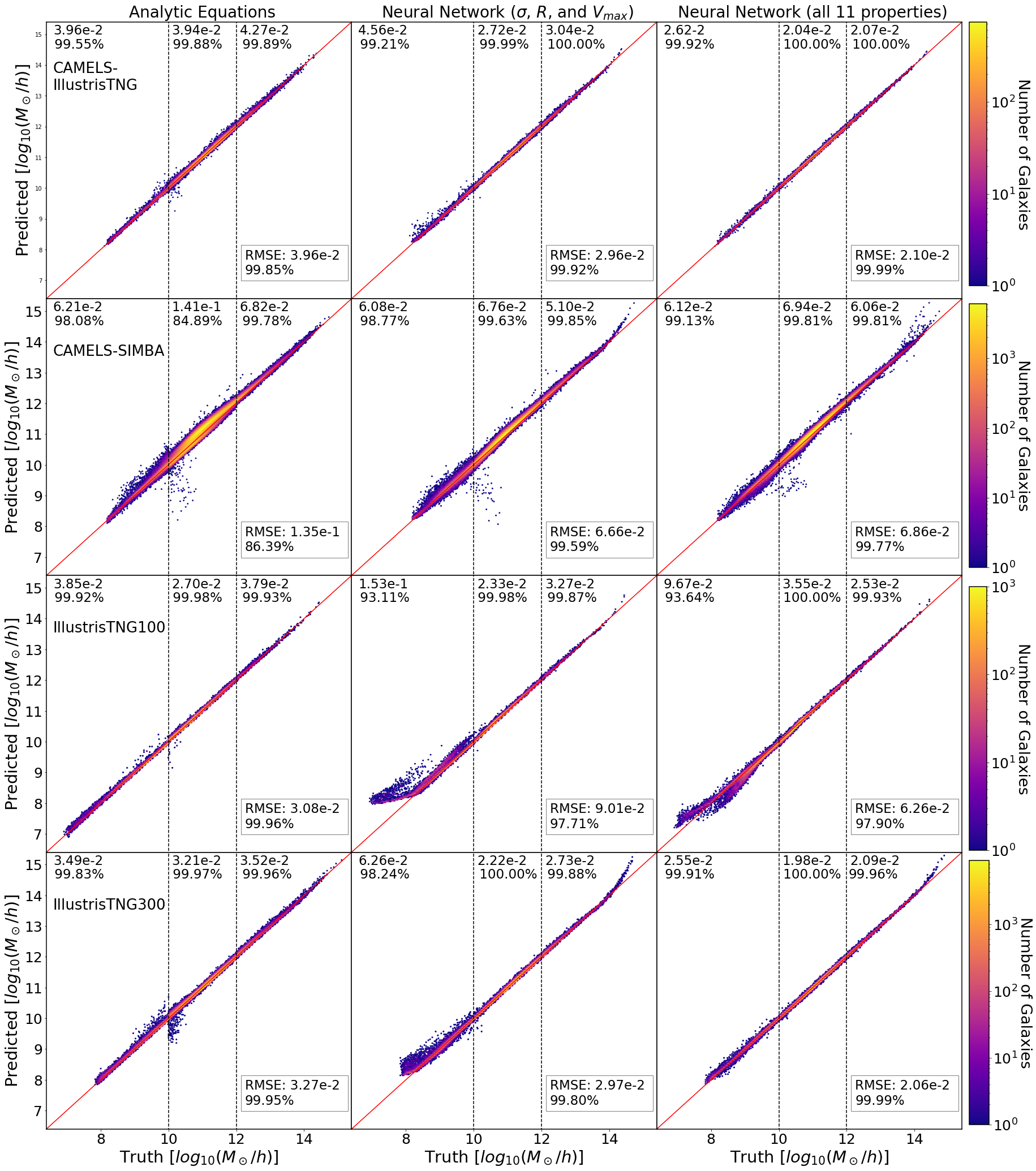}
    \caption{We fit analytic models to the CAMELS-IllustrisTNG simulations for three different mass ranges and test them on each simulation. This figure compares them to the results of the neural networks. From top to bottom, the rows illustrate the predictions for CAMELS-IllustrisTNG, CAMELS-SIMBA, IllustrisTNG100, and IllustrisTNG300 simulations. The first column shows the predictions from the analytic model shown in Equation \ref{Eq:formula}; the next column shows the predictions from the neural network trained using only three properties (${\sigma}$, $V_{\rm max}$, and $R$); and the last column shows the predictions from the neural network trained using all eleven subhalo properties. Moreover, the dotted lines at $10^{10} $ \(h^{-1}M_\odot\) and $10^{12}$ \(h^{-1}M_\odot\) divide each plot into three sections to indicate the mass ranges that the symbolic models were tested and trained on. The values quoted at the top of each plot for each mass range are the root mean square errors and the percentage of predictions within 0.2 dex of the truth, respectively. These values are also quoted at the bottom left of each plot for all the mass ranges of that simulation. As shown, in the high and low mass ends, the neural networks perform worse than the analytic models when tested on simulations other than CAMELS-IllustrisTNG.}
    \label{fig:all_models}
\end{figure*}

In Fig. \ref{fig:all_models} we compare the accuracy of these equations versus the one from two networks: one trained only using $R$, $\sigma$, and $V_{\rm max}$ (the variables present in the analytic formula) and the other trained using all properties. In all cases, the training has been carried out using subhalos of the CAMELS-IllustrisTNG simulations at $z=0$. We find that these equations are able to accurately predict the total mass of a subhalo from its radius, maximum circular velocity, and velocity dispersion, independent of whether the subhalo is from one simulation or another.

There are several interesting features to comment on. First, in general, the accuracy of the equation is worse than the network trained on the same three variables. This indicates that our formula is not capturing the full functional form learned by the network. Second, the model trained using all variables is, in general, more accurate than both the model trained with three variables and the analytic equation. This means that our formula does not account for the additional information on $M_{\rm tot}$ carried out by variables other than $R$, $V_{\rm max}$, and $\sigma$. Third, our equation is able to predict the total mass of subhalos with very low or very high total mass (beyond the range the model was trained on) more accurately than the neural networks. This may be happening because in that regime, the network is in extrapolation mode and may be subject to a number of factors limiting its accuracy, such as learning uninformative priors \citep{Villaescusa-Navarro_2020c}.

We then test the accuracy of Eq. \ref{Eq:formula} on subhalos from the different simulations at higher redshifts, finding the accuracy of the equation at those redshifts to be very similar as at $z=0$. We note that when using Eq. \ref{Eq:formula} at redshifts other than zero, we need to use the proper radius instead of the comoving radius (see appendix \ref{sec:variables})

Finally, we also test Eq. \ref{Eq:formula} on dark matter subhalos from CAMELS-Nbody simulations at redshifts $z = 0$ and $z=0.5$. We find that the accuracy of the analytic model is similar to the accuracy of the neural network that was trained using the same three variables (see Figure \ref{fig:N-Body}).

\begin{figure}
    \includegraphics[width=0.47\textwidth]{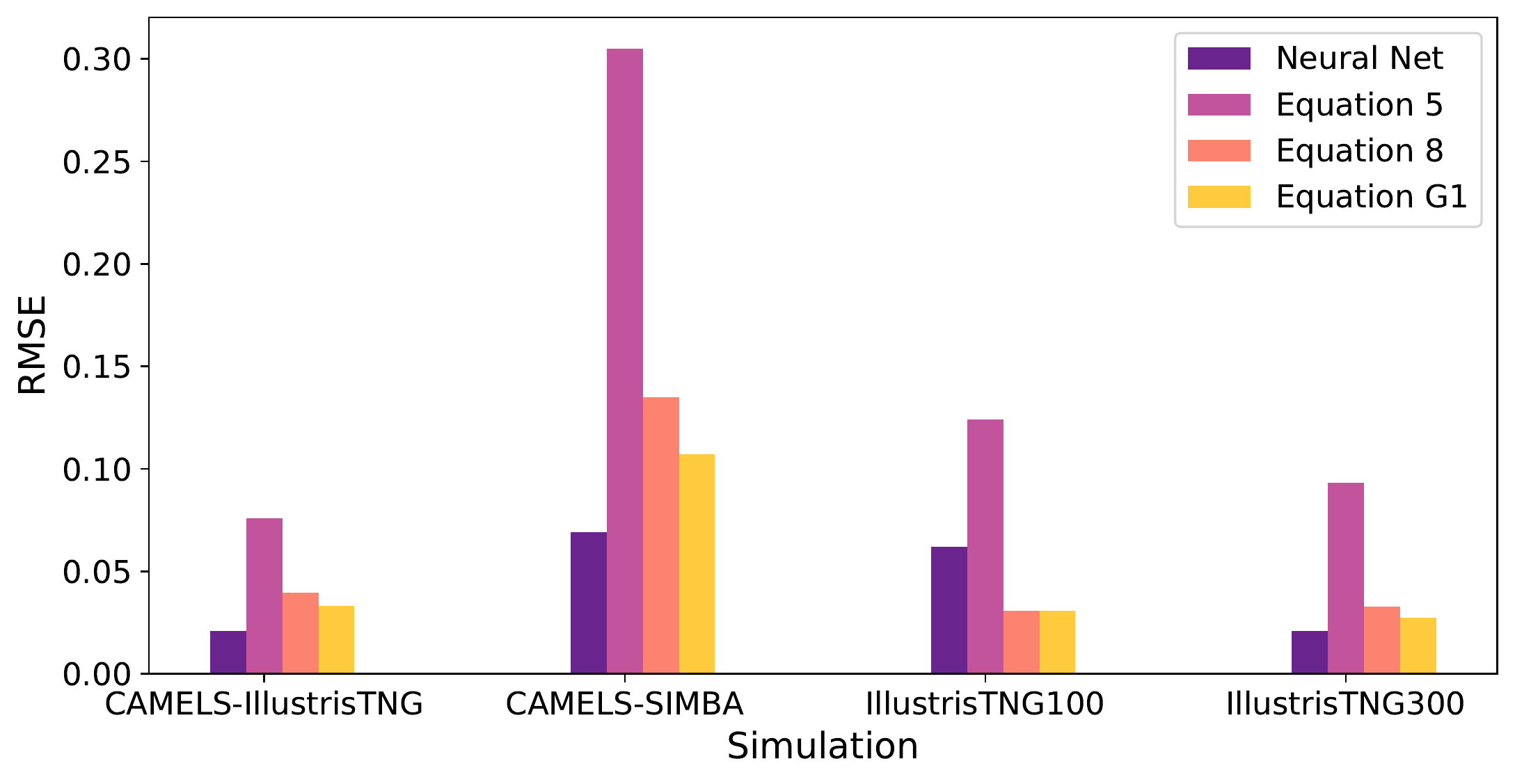}
    \caption{This barplot compares the root mean square errors of the various models when tested on the four simulations: CAMELS-IllustrisTNG, CAMELS-SIMBA, IllustrisTNG100, and IllustrisTNG300. The legend indicates the model associated with each color: Equations \ref{Eq:simple_equation}, \ref{Eq:simple_equation2}, \ref{Eq:formula}, and \ref{Eq:eureqa}. It can be seen that while the neural network model exhibits very high accuracy in the CAMELS-IllustrisTNG and IllustrisTNG300 simulations, it is unable to extrapolate well to CAMELS-SIMBA and IllustrisTNG100. Moreover, Equation \ref{Eq:simple_equation}, which is similar in form to the virial theorem, performs poorly on all simulations. In general, Equations \ref{Eq:formula} and \ref{Eq:eureqa} are the best analytic models in terms of accuracy and extrapolation ability, and exhibit similar RMSE.}
    \label{fig:barplot}
\end{figure}

Overall, our relatively simple equations are able to capture the main trend of the relation between the total mass of a subhalo and its other internal properties. We note that all above equations have been derived from the \textsc{PySR} software. In Appendix \ref{sec:eureqa} we present a set of equations found using the symbolic regression package \textsc{Eureqa} that performs with slightly higher accuracy than the above equations but have a more complex form. The root mean square errors for these equations are also included in the last row of Table \ref{tab:equations_table} for comparison. In Fig. \ref{fig:barplot} we show the RMSE values for the different simulations when using the neural networks or the different equations.

\section{Summary and Discussion}
\label{sec:conclusions}

The formation and evolution of galaxies is driven by numerous complex physical mechanisms whose details we do not yet fully understand. One possibility to improve our understanding of these processes is to find relationships between subhalo and galaxy properties that may reflect the underlying physics governing their formation and evolution. In this work we follow a generic methodology to search for one such relation connecting the total mass of a subhalo ($M_{\rm tot}$) with other internal properties ($\vec{\theta}$): $M_{\rm tot}=f(\vec{\theta})$. We have used machine learning methods to approximate the function $f$ using state-of-the-art hydrodynamic simulations and learned about its properties.

The most important findings of this work are:
\begin{itemize}
\item Neural networks are able to find a relation between
the total mass of a subhalo and its other properties, independently of the subhalo type (i.e. a central or a satellite). The model is able to accurately predict the total mass of subhalos from simulations with different cosmologies, different astrophysical models, different subgrid models and hydrodynamic solvers, different volumes, different resolutions, and different redshifts than the ones it has been trained on. The surprising accuracy achieved by the network in these different extrapolations indicates that the network may have found a universal relationship within the subhalo properties.
\item Our networks only work at redshifts $z>0$ if the proper radius is used instead of the comoving radius. This indicates that the learned relation depends on physical variables and not comoving ones. 
\item We find that the models trained on fewer subhalo variables ($R$, $V_{\rm max}$, $\sigma$, $M_g$, and ${\rm SFR}$) are able to extrapolate slightly better than models trained using all subhalo properties. This could be due to the fact that the network may be leveraging information that is only present in some simulations, and therefore it extrapolates worse when those conditions are no longer met.
\item When the network is trained using $R$, $V_{\rm max}$, and $\sigma$ from subhalos, it is also able to predict the total mass of subhalos from N-body simulations, although with higher scatter for subhalos with low mass.
\item Our results are robust to the type of simulation used for training the networks, including very different subgrid galaxy formation models.
\item A simple relation of the form $M_{\rm tot}=AR^\alpha \sigma^\beta$ is able to roughly capture the main relation between the total mass of a subhalo and its other properties. However, it exhibits significant biases in the low and high mass ends. We find that including $V_{\rm max}$ is crucial to properly capturing the underlying relation in the low mass end.
\item Eq. \ref{Eq:formula} is a much more accurate expression that works for subhalos from simulations with different cosmologies, astrophysics, subgrid physics, volumes, resolutions, and at different redshifts. Furthermore, this equation is able to extrapolate in the low and high mass end more precisely than the neural networks. 
\item The neural networks achieve a higher accuracy than Eq. \ref{Eq:formula}, indicating that: 1) our expression may only be an approximation to the relation found by the network, and 2) that relation should also contain a dependence on other variables, such as gas mass and star-formation rate, that our equation lacks.
\end{itemize}

The rather surprising extrapolation properties of both our neural networks and the analytic equation may indicate that the relation between the total mass of a subhalo and its other properties is set by a relatively simple, universal, law. 

To our knowledge, the equations derived in this work are new, and while complex, they are able to capture the main trend of the relation between the mass and other subhalo properties. These equations may help us in understanding the underlying physics behind subhalo/galaxy formation and evolution that, ultimately, is responsible for shaping the distribution of dark matter, gas, and stars in the Universe. 

We however believe that simpler and more accurate equations that capture the apparent universal relationship between the total mass of a subhalo and its other internal properties can be found. We leave this for future work.

We also think that although these results are generic, their details may depend on the particular algorithm used to identify subhalos and halos in the simulations. For instance, a different criteria in the definition of radius or velocity dispersion may led to subhalos for which our network and analytic expression do not work. However, in this case we believe retraining the network and refitting the best values of Eq. \ref{Eq:formula} may fix this issue.

Lastly, we attempt to provide a physical interpretation to our results. From Eq. \ref{Eq:simple_equation} we may guess that the network is learning some version of the virial theorem, that states that in a self-gravitating body the total kinetic ($T$) energy relates to the gravitational potential energy ($U$) via $2T+U=0$, or $M=V^2R/G$. The fact that the relation learned by the network depends on the proper (physical) size of subhalos, instead of their comoving scale, reinforces this point. 

However, it cannot be that same relation because the variables used do not correspond with the ones appearing in the virial theorem. E.g. our radius, $R$ is defined as the radius containing half of the subhalo total mass, while in the virial theorem $R$ is the radius of the virial system. Besides, our network and analytic expression make use of $V_{\rm max}$, that does not appear in the virial theorem. On top of this, we know that variables such as the gas mass and the star-formation rate are being used by the network to improve its accuracy. However, the network may be using all these variables as a proxy for the ones appearing in the virial theorem. 

The universality of our results could also be explained if the underlying relation arises from the virial theorem, that is derived from very generic arguments related to gravity, and therefore are not affected by cosmology, astrophysics, simulation volume, simulation resolution, subgrid physics, and redshift. 

Overall, this paper presents a generic formalism to search for relationships in high-dimensional spaces and approximate them with analytic expressions. In future work we will use this formalism to search for relations in galaxy properties from actual observations and derive analytic equations from them.

\section*{ACKNOWLEDGEMENTS}
We thank Rachel Somerville for useful conversations. This work has made use of the Tiger cluster of Princeton University and the Iron and Popeye clusters at the Flatiron Institute, which is supported by the Simons Foundation. FVN acknowledges funding from the WFIRST program through NNG26PJ30C, NNN12AA01C, and from NSF through AST-2108078. DAA was supported in part by NSF grants AST-2009687 and AST-2108944.
Details on the CAMELS simulations can be found in \url{https://www.camel-simulations.org}. 

\appendix

\section{Central vs. Satellite Galaxies}
\label{sec:centrals_satellites}

\begin{figure}
    \centering
    \includegraphics[width=1\textwidth]{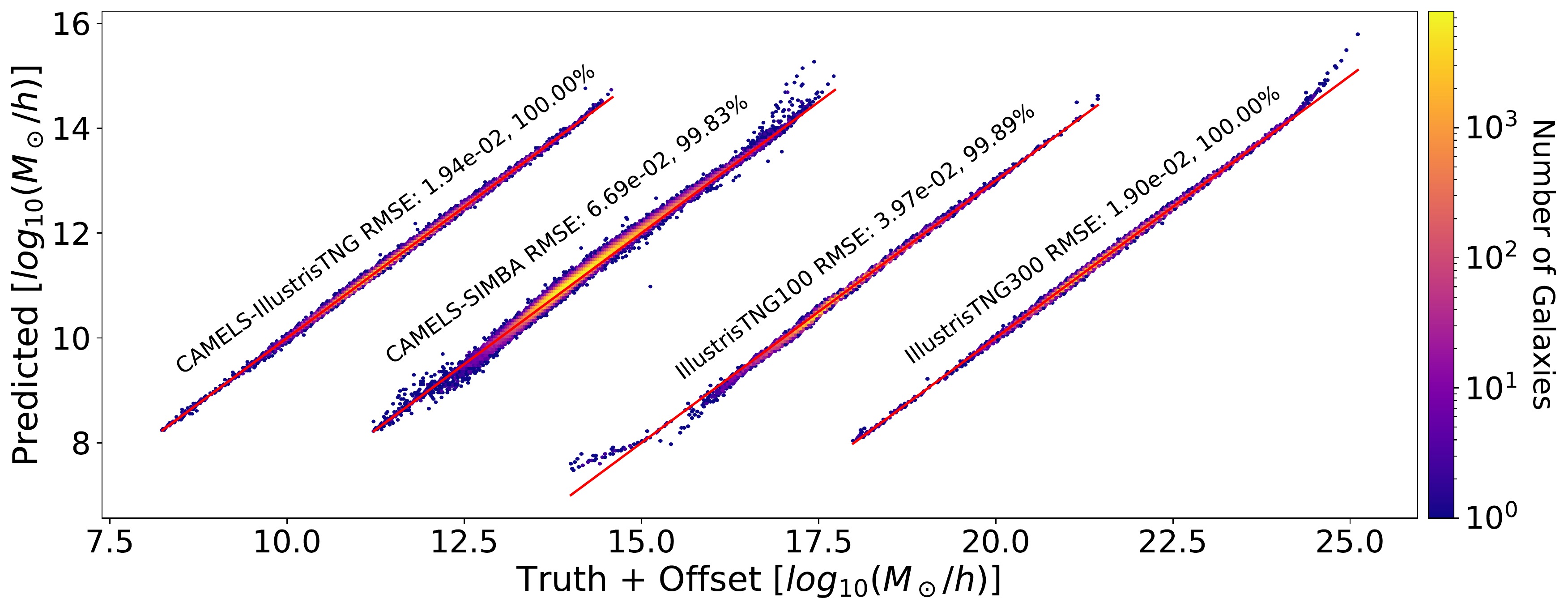}
    \caption{We test the neural network trained to predict the total subhalo mass from eleven other subhalo properties on the central subhalos from the CAMELS-IllustrisTNG, CAMELS-SIMBA, IllustrisTNG100, and IllustrisTNG300 simulations. In this figure, we plot the model predictions against the truth for each simulation on one set of axes to conserve space. For the CAMELS-IllustisTNG subhalos, we plot the predicted mass against the truth. For the CAMELS-SIMBA subhalos, we plot the predicted mass against the truth plus 3 dex. For the IllustrisTNG100 subhalos, we plot the predicted mass against the truth plus 7 dex. Finally, for the IllustrisTNG300 subhalos, we plot the predicted mass against the truth plus 10 dex. As can be seen, the model is able to perform with very high accuracy for the central subhalos, as more than 99\% of its predictions lie within 0.2 dex of the truth for all simulations.}
    \label{fig:centrals}
\end{figure}

\begin{figure}[h]
    \centering
    \includegraphics[width=1\textwidth]{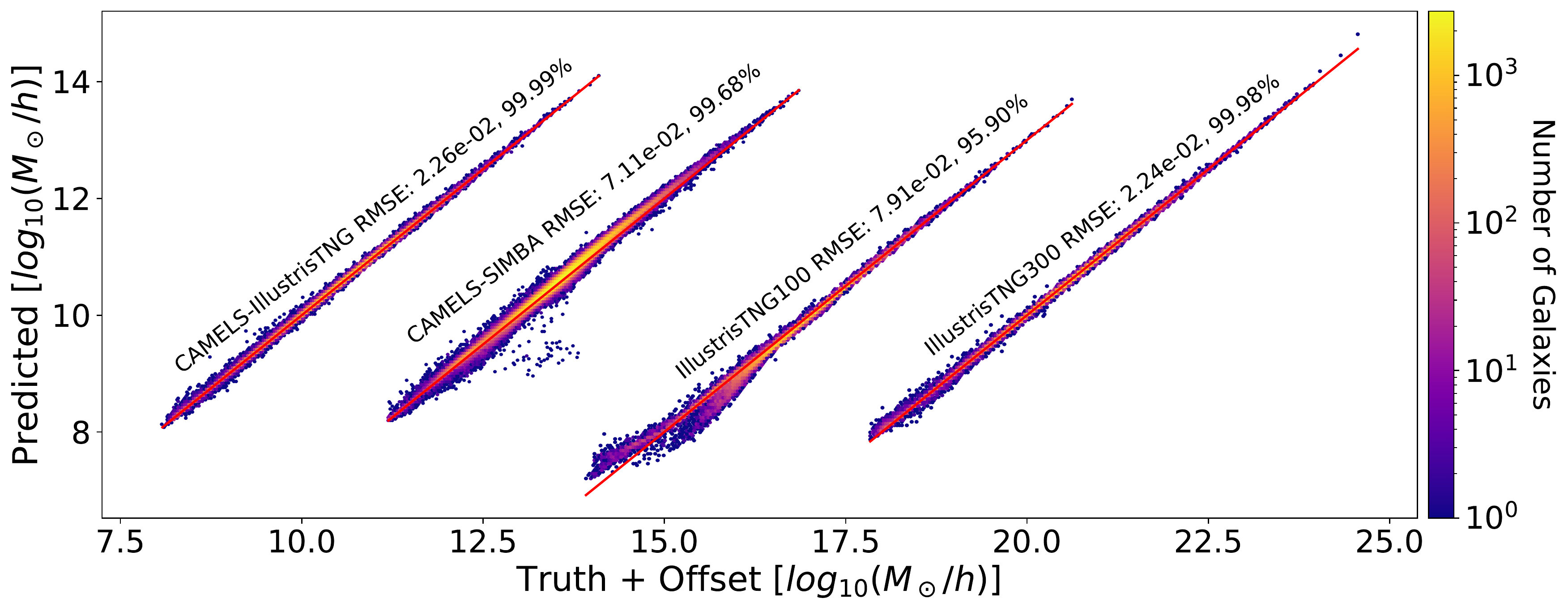}
    \caption{Same as Fig. \ref{fig:centrals} but for the subhalo satellites from the four different simulations.}
    \label{fig:satellites}
\end{figure}

In this section, we compare the neural network's predictions when subhalos are split into central and satellites for all different simulation types. The results for the central subhalos are depicted in Figure \ref{fig:centrals} while the results for the satellites are shown in Figure \ref{fig:satellites}. As can be seen, while the latter exhibits slightly higher error, the model is still able to accurately predict the total mass for both types of subhalos. This is a surprising result given that the neural network was not given information that directly differentiates the satellites from the centrals, which may indicate that such information is implicitly encoded in relations between the input subhalo properties. Moreover, there are several interesting points to note. First, it is known that satellites with eccentric orbits are being tidally stripped in a time dependent manner, as well as tidally shocked, as they reach pericenter. Hence, it is not obvious why the relation found by the neural network, one we believe to be rooted in the virial theorem, holds in this case where the system may not be virialized. A possible explanation is that subhalos that are undergoing mass stripping may still be in a semi-equilibrium state, which allows the conditions of the virial theorem to apply. However, further investigation on such satellites is needed to determine their effect on the neural network's predictions. Second, while the neural network does not explicitly account for the tidal radius of the satellites or their location relative to their hosts, which are essential pieces of information for the orbiting bodies, it is still able to accurately predict the total subhalo mass. Here, we believe that the radius used to train the neural network implicitly encodes the information about the tidal radius of the satellites. Given that the radius we use is defined to contain half the total subhalo mass, the effect of tidal stripping would reduce both mass and radius. Third, it is possible that the larger scatter seen for satellites is a result of both such physical effects and numerical effects related to the specific substructure-finding algorithm, which may sometimes have difficulty identifying the full extent of subhalos when they pass close to the host's center. A more detailed analysis of the relative dynamics between the satellites and their hosts in the context of the found relations is needed, and a more elaborate treatment of this can possibly reduce the scatter in the fittings. We leave this for future investigation.

\section{Higher Redshifts}
\label{sec:redshift}

\begin{figure*}[ht]
\gridline{\fig{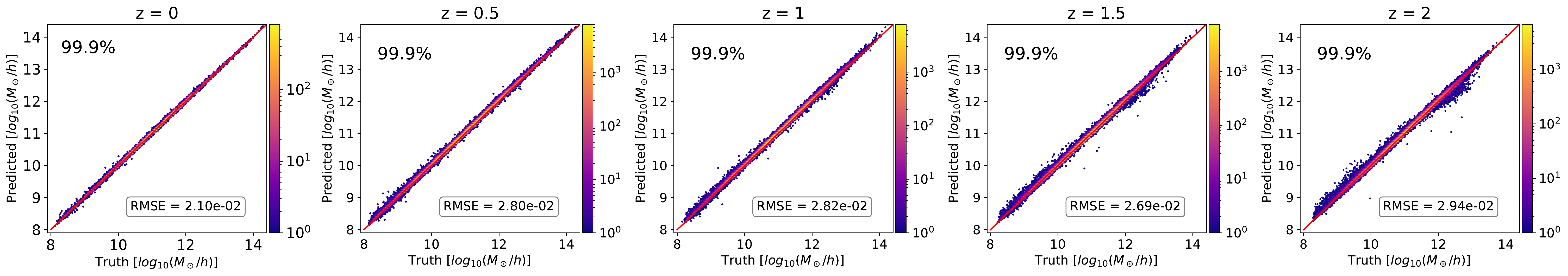}{1\textwidth}{CAMELS-IllustrisTNG}}
\gridline{\fig{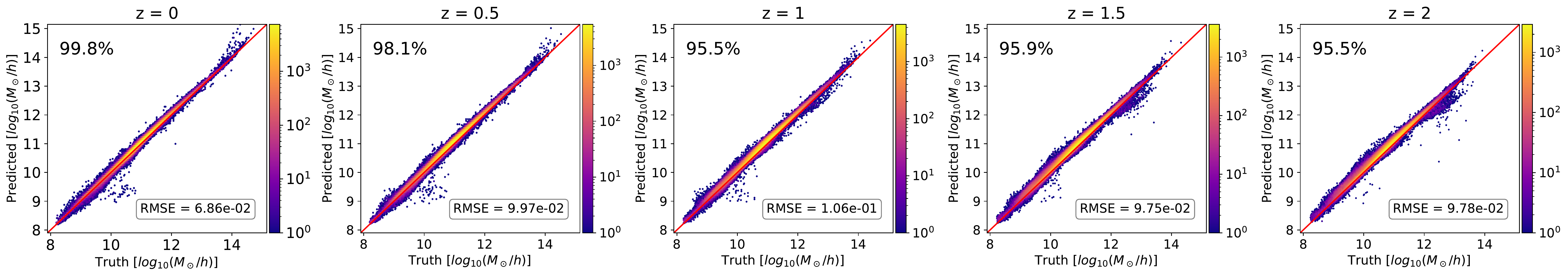}{1\textwidth}{CAMELS-SIMBA}}
\gridline{\fig{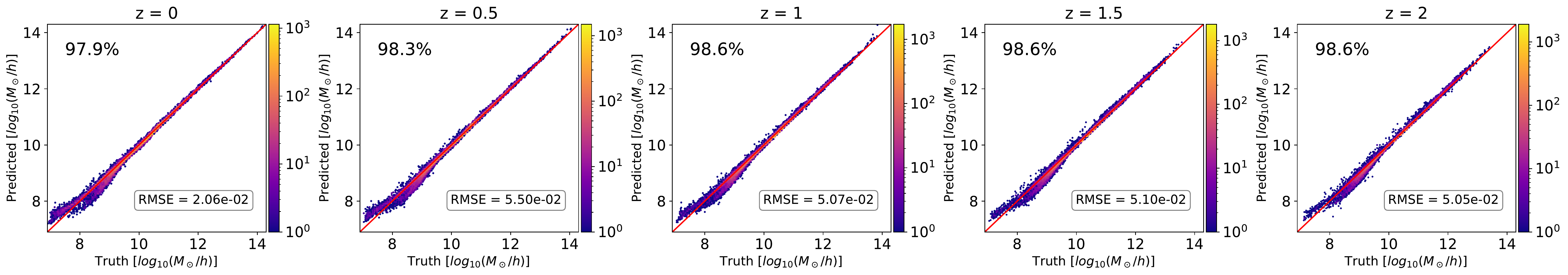}{1\textwidth}{IllustrisTNG100}}
\gridline{\fig{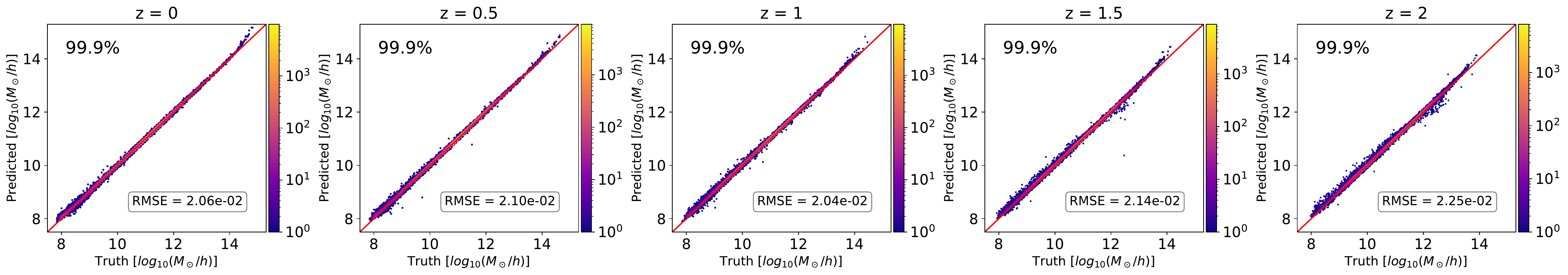}{1\textwidth}{IllustrisTNG300}}
\caption{We test the neural network that was trained on CAMELS-IllustrisTNG subhalos using all 11 properties at $z=0$ on subhalos from different simulations at different redshifts. The percentage of predictions that fall within 0.2 dex of the true value is quoted at the top left of each plot, along with the root mean square error at the bottom right. As demonstrated, the model is able to extrapolate in redshift very accurately in all cases. This may indicate that the network has indeed found a universal relationship among subhalo properties. 
\label{fig:redshifts}}
\end{figure*}

Here we quantify how well our neural network, trained on CAMELS-IllustrisTNG subhalos at $z=0$, can predict the total mass of subhalos from the different simulations at higher redshifts. We show in Figure \ref{fig:redshifts} the results of testing the network on subhalos at redshifts $z=0$, $z=0.5$, $z=1$, $z=1.5$, and $z=2$. We find that our model can accurately predict the total mass of the subhalos at all considered redshifts, independently of the simulations they come from. We emphasize that CAMELS-SIMBA simulations do not use the same hydrodynamic code as the IllustrisTNG-like simulations, and utilizes a very distinct subgrid physics model. Therefore, the high accuracy of the model's predictions at higher redshifts for the SIMBA subhalos is not a trivial result. This reinforces our belief that the neural network may have found a universal relation since it has never seen subhalos at redshifts other than $z=0$. 

It is important to emphasize that in this exercise we have fed the network with the proper radius of a subhalo instead of its comoving radius. In Appendix \ref{sec:variables} we show what happens when the comoving radius is used at higher redshifts. This indicates that the relation learned by the network depends on the physical radius and not on the comoving radius. 

\section{Comoving versus Proper Radius}
\label{sec:variables}

\begin{figure}[h]
    \centering
    \includegraphics[width=0.82\textwidth]{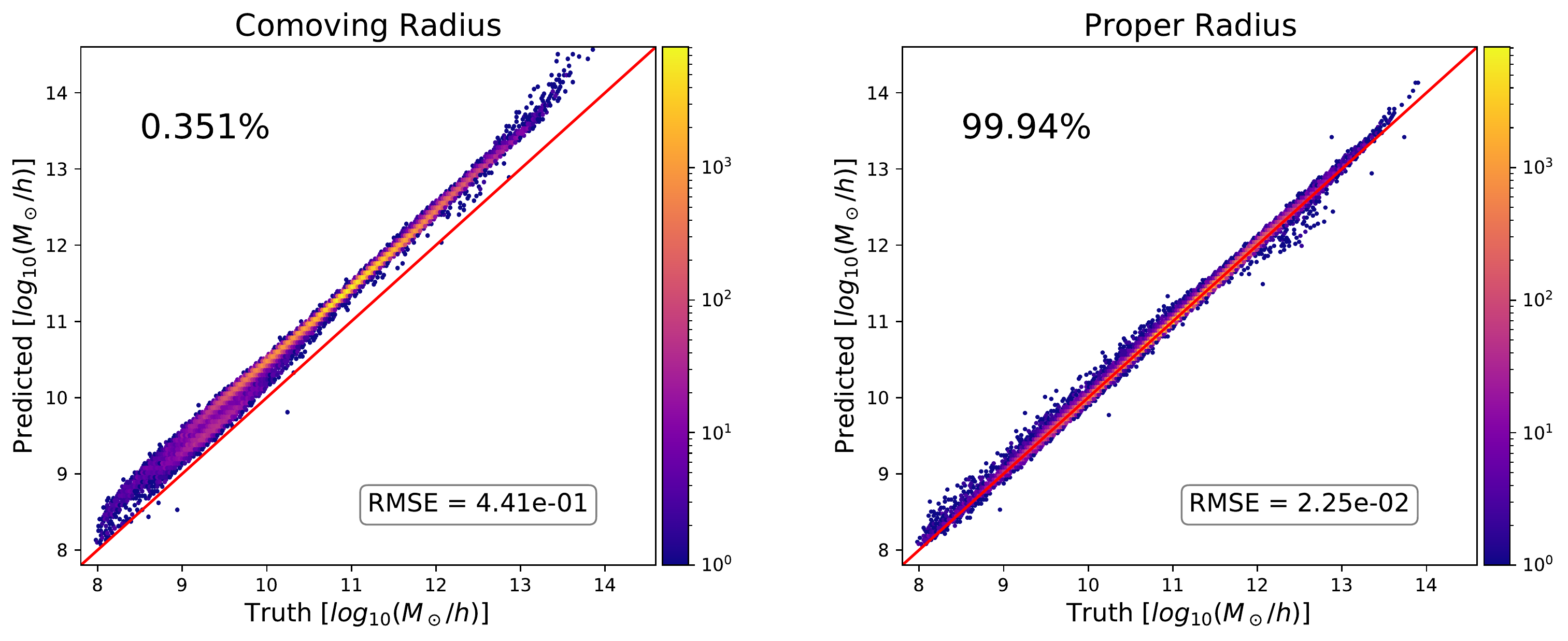}
    \caption{These plots show the large difference in accuracy of the neural network's predictions when using the comoving radius (left) versus the proper  radius (right) of the subhalos at redshift $z = 2$. Here, the neural network was trained on subhalos of the CAMELS-IllustrisTNG simulations at $z = 0$ and tested on subhalos from the IllustrisTNG300 simulations at $z = 2$. The root mean square error is quoted in the bottom right and the percentage of predictions that lie within 0.2 dex of the actual value is quoted on the top left. This shows that the relation learned by the network depends on proper lengths and not comoving ones.}
    \label{fig:comoving_proper}
\end{figure}

While exploring the neural network's performance at higher redshifts, we found that using different definitions of the subhalo radius led to drastically different results. Namely, using the comoving radius, defined as the proper (physical) radius divided by the scale factor $a=1/(1+z)$, for subhalos at redshifts higher than z = 0 caused the neural network predictions to have a vertical shift, resulting in very low accuracy. This can be seen in the left panel of Fig. \ref{fig:comoving_proper}. We note that since the network was trained at $z=0$, where the comoving radius equals the proper radius, it is not obvious which radius definition should be used. Thus, we tested the network using the proper radius of the subhalos and show the results in the right panel of Fig. \ref{fig:comoving_proper}. As can be seen, in this case the network accuracy reaches similar values as the model at $z=0$. We thus conclude that the relation learned by the network, and the one that the network employs to extrapolate to other redshifts and simulations, uses the proper radius of the subhalos and not their comoving one.

\begin{figure}[h!]
    \centering
    \includegraphics[width=1.03\textwidth]{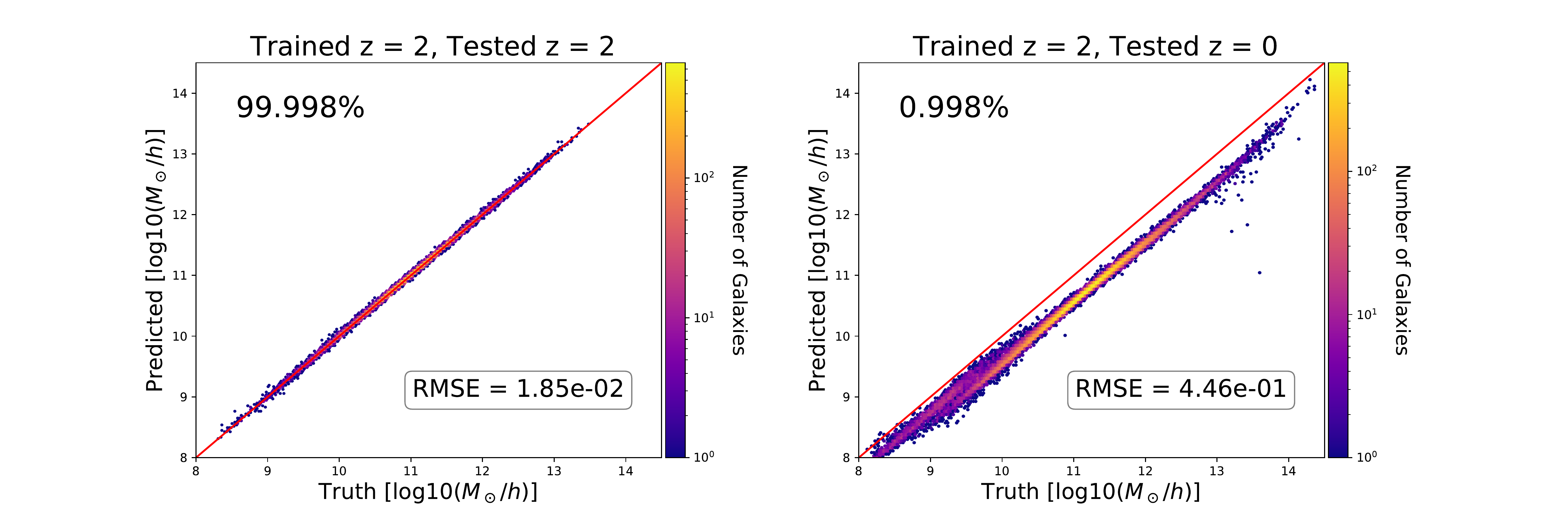}
    \caption{We trained a neural network to learn the total subhalo mass from the eleven subhalo properties of the CAMELS-IllustrisTNG simulations at redshift $z=2$ using the comoving radius as a variable, and tested the model on the subhalos from CAMELS-IllustrisTNG at redshift $z=0$. At this redshift, the comoving and proper radius definitions are equal. As shown, the model performs poorly when predicting the total mass for the subhalos at a different redshift from the subhalos used during training. This illustrates the importance of using the proper variables when training the network if extrapolation properties are desired. The root mean square error is quoted in the bottom right and the percentage of predictions that lie within 0.2 dex of the actual value is quoted on the top left.}
    \label{fig:z2_z0}
\end{figure}

This is actually a very important point that shows the importance of training neural networks using the proper variables if extrapolation properties are desired. To illustrate this point, we train a neural network using subhalos of the CAMELS-IllustrisTNG simulations at $z=2$ using the comoving radius as one of the input variables. The results of testing the network at $z=2$ are shown on the left panel of Fig. \ref{fig:z2_z0}. It can be seen that the network achieves a very high accuracy when predicting the total mass of subhalos at $z=2$. The right panel of Fig. \ref{fig:z2_z0} shows instead what happens when the network is tested on subhalos at $z=0$. We note that at this redshift, the proper and comoving definitions are equal. Evidently, the network is not able to predict the correct total mass of the subhalos. This case shows how training the neural network on variables that are not the ones behind a relation may significantly affect the extrapolation properties of the network itself. It is however important to emphasize that the proper variables behind a given case may not be known a-priori.

\section{N-body Simulations}
\label{sec:N-Body}

\begin{figure}[h]
    \centering
    \includegraphics[width=1\textwidth]{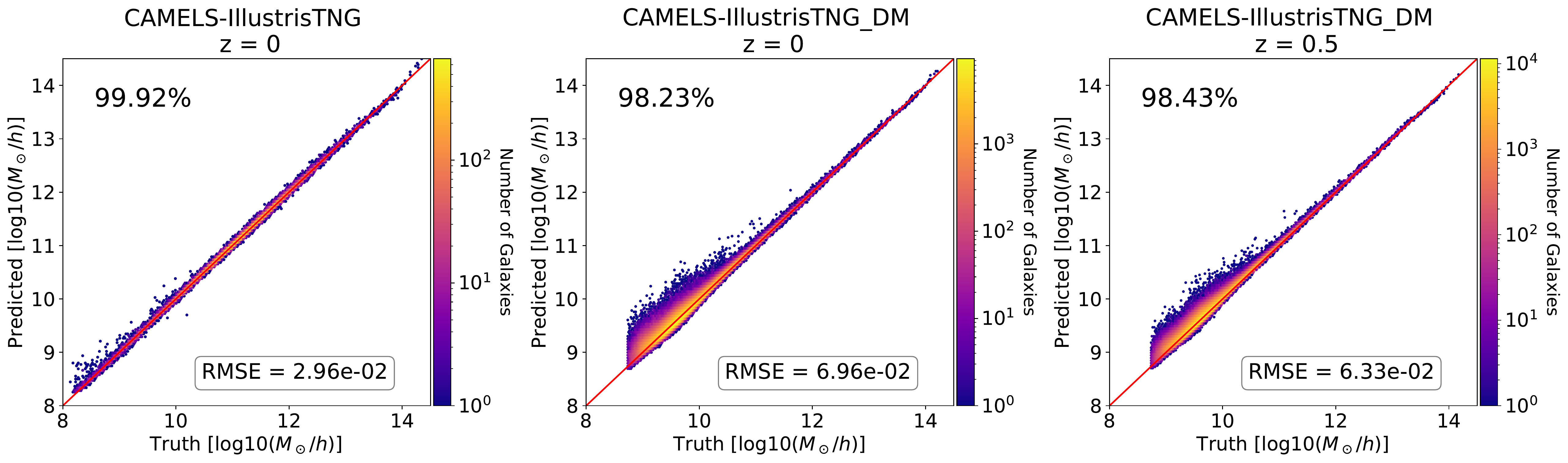}
    \caption{We trained a neural network model to predict the subhalo total mass from $R$, ${\sigma}$, and $V_{max}$ using the subhalos from the CAMELS-IllustrisTNG suite (at z=0). The leftmost plot shows the model predictions for a test set of the  CAMELS-IllustrisTNG simulations (at z=0). The following two plots depict the model predictions when tested on the N-body subhalos from the CAMELS-IllustrisTNG Dark Matter simulations at redshifts z = 0 and 0.5. The root mean square error is quoted in the bottom right and the percentage of predictions that lie within 0.2 dex of the actual value is quoted on the top left.}
    \label{fig:N-Body}
\end{figure}

Here we study if the model that is trained on subhalos from hydrodynamic simulations can predict the total mass of subhalos from gravity-only N-body simulations. Since the subhalos of the N-body simulations do not have some properties that are only present in the subhalos of the hydrodynamic simulations, such as SFR, gas mass, and others, we use the model that was trained on the three variables the subhalos from both simulations share: $R$, $V_{\rm max}$, and $\sigma$. The left panel of Fig. \ref{fig:N-Body} shows the results when the model is tested on subhalos from the CAMELS-IllustrisTNG simulations at $z=0$. As can be seen, the model is able to accurately predict the total mass of these subhalos over the whole mass spectrum. The middle and right panels of that figure show the results when that model is tested on subhalos from N-body simulations with different cosmologies at redshifts z=0 and z=0.5. Here, we also find that the model is able to predict the total mass of the N-body subhalos with high accuracy, although the RMSE is slightly higher than that for subhalos of hydrodynamic simulations and the fraction of N-body subhalos with its total mass within 0.2 dex of their true value is slightly lower than that for the hydrodynamic simulations. 

These results show that the relation found by the network is not unique to subhalos from hydrodynamic simulations, but points towards something more fundamental that may just depend on gravity itself, like some version of the virial theorem. 

\section{Training on Other Simulations} \label{sec:other_sims}

\begin{figure}
    \centering
    \includegraphics[width=1\textwidth]{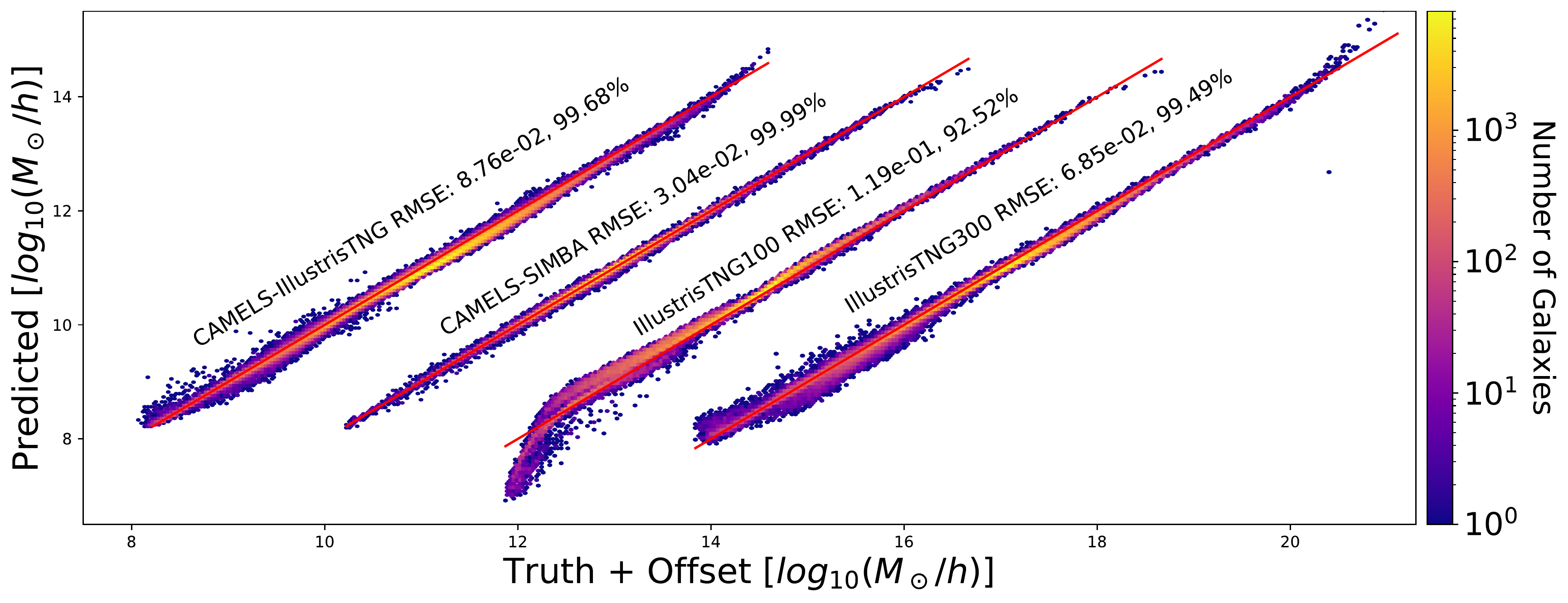}
    \caption{We train a neural network to learn the total subhalo mass frrom eleven other subhalo properties on the CAMELS-SIMBA simulations at redshift $z=0$. We then test the model on a test set of subhalos from the CAMELS-SIMBA simulations, as well as on the subhalos of the CAMELS-IllustrisTNG, IllustrisTNG100, and IllustrisTNG300 simulations. In this figure, we plot the model predictions against the truth for each simulation on one set of axes to conserve space. For the CAMELS-IllustisTNG subhalos, we plot the predicted mass against the truth. For the CAMELS-SIMBA subhalos, we plot the predicted mass against the truth plus 2 dex. For the IllustrisTNG100 subhalos, we plot the predicted mass against the truth plus 4 dex. Finally, for the IllustrisTNG300 subhalos, we plot the predicted mass against the truth plus 6 dex. It is apparent that the model is able to predict with relatively high accuracy as more than 92\% of the predictions lie within 0.2 dex of the truth for all simulations. However, compared to the neural network that was trained on the CAMELS-IllustrisTNG subhalos (see Figure \ref{fig:nn_all}), this model is performing worse for all simulations except CAMELS-SIMBA. The root mean square error and the percentage of the predictions within 0.2 dex of the true value are quoted next to each scatter plot in the figure.}
    \label{fig: simba_train}
\end{figure}

In order to quantify the dependence of our neural network results on their training set, we train two networks using subhalos from the CAMELS-SIMBA and IllustrisTNG300 simulations using all 11 properties. We then test each model on the subhalos of the other simulations. We show the results in Fig. \ref{fig: simba_train} for the network trained on subhalos from the CAMELS-SIMBA simulations. It can be seen that even though the model is using a robust relation to predict the total mass, it is not as accurate as the model that was trained on subhalos from the CAMELS-IllustrisTNG simulations. The RMSE values for the neural network trained on CAMELS-SIMBA are: $7.67\times10^{-3}$ (CAMELS-IllustrisTNG), $9.27\times10^{-4}$ (CAMELS-SIMBA), $1.41\times10^{-2}$ (IllustrisTNG100), and $4.70\times10^{-3}$ (IllustrisTNG300). 

\begin{figure}
    \centering
    \includegraphics[width=1\textwidth]{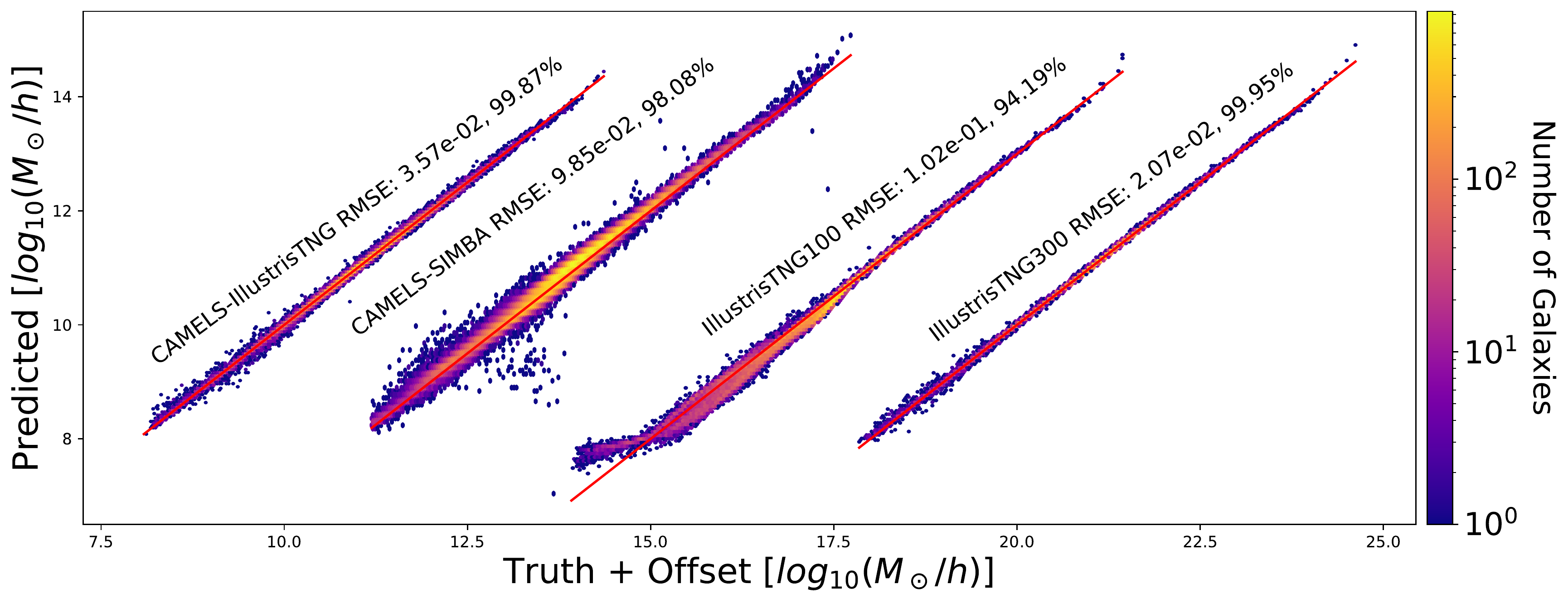}
    \caption{Same as Fig. \ref{fig: simba_train} but training on subhalos from the IllustrisTNG300 simulation.}
    \label{fig:tng300_train}
\end{figure}

In Fig. \ref{fig:tng300_train} we show instead the results of the network trained on subhalos of the IllustrisTNG300 simulation. The accuracy of these predictions are similar to those of the model trained on CAMELS-SIMBA. The RMSE values for the neural network trained on IllusrisTNG300 are: $3.56\times10^{-2}$ (CAMELS-IllustrisTNG), $9.84\times10^{-2}$ (CAMELS-SIMBA), $1.14\times10^{-1}$ (IllustrisTNG100), and $2.06\times10^{-2}$ (IllustrisTNG300). 

These results show that although the data used in the training set can have a relatively minor effect on the accuracy of the network, for which we provide further interpretation in Section \ref{subsec:nn}, the main conclusions are unchanged, i.e. the network can accurately predict the total mass of subhalos from simulations with different cosmologies, astrophysics, subgrid physics, resolutions, and volumes than the ones used to train the network.

\section{Input Properties}
\begin{figure}
    \centering
    \includegraphics[width=1\textwidth]{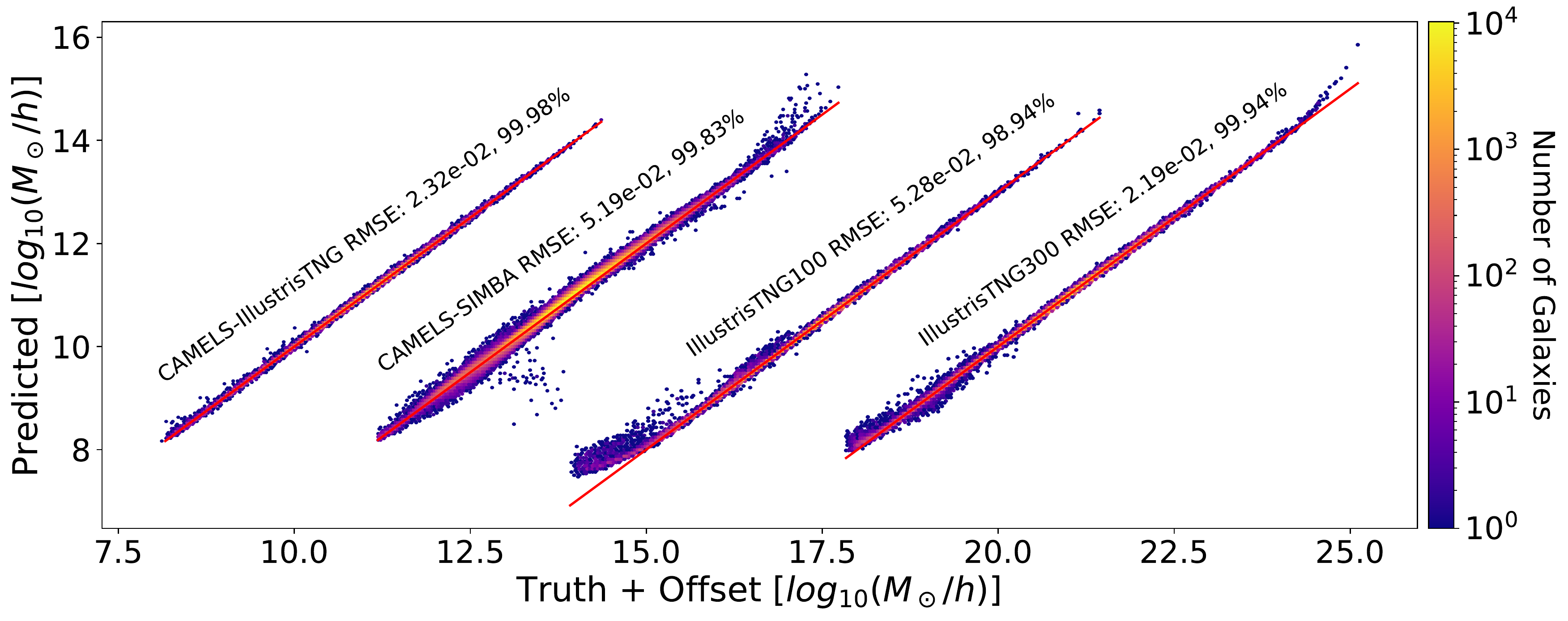}
    \caption{We train a neural network using the CAMELS IllustrisTNG simulations to learn the total subhalo mass from five selected properties: $R$, ${\sigma}$, $V_{\rm max}$, $SFR$, $M_{\rm g}$ at $z=0$. In this figure, we plot the model predictions against the truth for each simulation on one set of axes to conserve space. For the CAMELS-IllustisTNG subhalos, we plot the predicted against the truth.  For the CAMELS-SIMBA subhalos, we plot the predicted against the truth plus 3 dex. For the IllustrisTNG100 subhalos, we plot the predicted against the truth plus 7 dex. Finally, for the IllustrisTNG300 subhalos, we plot the predicted against the truth plus 10 dex. The root mean square error and the percentage of the predictions within 0.2 dex of the true value are quoted above each scatter plot in the figure. We can see that even from only five properties, the neural network is able to predict the total mass with very low error. Even though it is performing worse than the original model for the CAMELS-IllustristTNG simulation, it is doing better for the IllustrisTNG100 and CAMELS-SIMBA subhalos.}
    \label{fig:nn_5vars}
\end{figure}

Here we investigate the effects of reducing the number of input subhalo properties on the neural network's ability to predict the total subhalo mass. For that, we train a neural network to learn the total mass using only five subhalo properties: $R$, ${\sigma}$, $V_{\rm max}$, ${\rm SFR}$, and $M_g$ on subhalos from the CAMELS-IllustrisTNG simulations. As illustrated in Figure \ref{fig:nn_5vars}, the neural network is still able to predict the total mass with high accuracy (see also Table \ref{tab:nn_table}). Notably, the model predictions for subhalos of the CAMELS-SIMBA and IllustrisTNG100 simulations exhibit lower root mean square errors compared to the predictions from the model trained on all eleven properties. 

A potential explanation for this behaviour is that when the training is carried out using all variables, the neural network may be exploiting relations that are unique to a particular simulation. These relations may be numerical artifacts or peculiarities of the particular simulations considered. In that case, training with more variables will improve the network precision when the network is tested on simulations of the same kind as the ones used for training, but may actually perform worse when tested on other simulations. By training on a smaller number of variables, the network may learn more robust correlations that can extrapolate better. 

We note however that the differences between the results obtained by using all variables and 5 variables is not very large, indicating that the original network is already marginalizing over uninformative variables.

\section{Eureqa Equations} \label{sec:eureqa}

We train symbolic regression models using the \textsc{Eureqa} package on the CAMELS-IllustrisTNG simulations to search for analytic expressions that can predict the total subhalo mass from the eleven other subhalo properties. First, we train a model on a subset of subhalos that span the entire mass range from the CAMELS-IllustrisTNG simulations. While the equation that the model discovered is able to capture the main trend of the total subhalo mass, it performs with lower accuracy than the analytic equations discussed in Section \ref{subsec:symbolic}. To improve the results and aid the model's search, we train symbolic regression models on subhalos spanning the three different mass ranges used in Section \ref{subsec:symbolic}: 1) $M_{\rm tot}< 10^{10}~h^{-1}M_\odot$, 2) $10^{10}~h^{-1}M_\odot < M_{\rm tot}<10^{12}~h^{-1}M_\odot$, 3) $M_{\rm tot}>10^{12}~h^{-1}M_\odot$. The equations that were found are:

\begin{equation}
    M_{\rm tot}= 
\begin{cases}
    10^{5.60}R^{2.72 + 2.00\log\tilde{\sigma} - 0.26\log\tilde{R}}V_{\rm max}^{1.65 - 0.68\log\tilde{\sigma} \log\tilde{R}}, & \text{if } M_{\rm tot}\leq 10^{10}~h^{-1}M_\odot\\
    
    10^{5.77}R^{15.48 + 3.82\log\tilde{V_{\rm max}} \log\tilde{\sigma}}V_{\rm max}^{1.89\log\tilde{V_{\rm max}}\log\tilde{R}- 3.93}\sigma^{-5.17 - 1.93\log\tilde{\sigma} \log\tilde{R}}, & \text{if } 10^{10}~h^{-1}M_\odot<M_{\rm tot}<10^{12}~h^{-1}M_\odot\\
    
    10^{9.74}R^{1.94 + 0.41\log ^2 \tilde{\sigma}}\sigma^{1.96 + \log\tilde{R} - 0.02\log ^2 \tilde{\sigma}}, & \text{if } M_{\rm tot}\geq 10^{12}~h^{-1}M_\odot\\
\end{cases}
\label{Eq:eureqa}
\end{equation}

In the above equations, we define $\tilde{R} = R/R_0$, $\tilde{\sigma} = \sigma/\sigma_0$, and $\tilde{V_{max}} = V_{\rm max}/V_{\rm max 0}$. For $M_{\rm tot}< 10^{10}~h^{-1}M_\odot$, $R_0 = 2.41$ $h^{-1}{\rm kpc}$ and $\sigma_0 = 25.6$ km/s. For $10^{10}~h^{-1}M_\odot < M_{\rm tot}<10^{12}~h^{-1}M_\odot$, $R_0 = 30.84$ $h^{-1}{\rm kpc}$, $\sigma_0 = 63.25$ km/s, and $V_{\rm max 0} = 123.33$ km/s. For $M_{\rm tot}>10^{12}~h^{-1}M_\odot$, $\sigma_0 = 148.96$ km/s and $R_0 = 113.24$ $h^{-1}{\rm kpc}$. We note that the units of $R$, $\sigma$, and $V_{\rm max}$ are $h^{-1}{\rm kpc}$, km/s, and km/s, respectively.

\begin{figure*}[h!]
    \centering
    \includegraphics[width=1\textwidth]{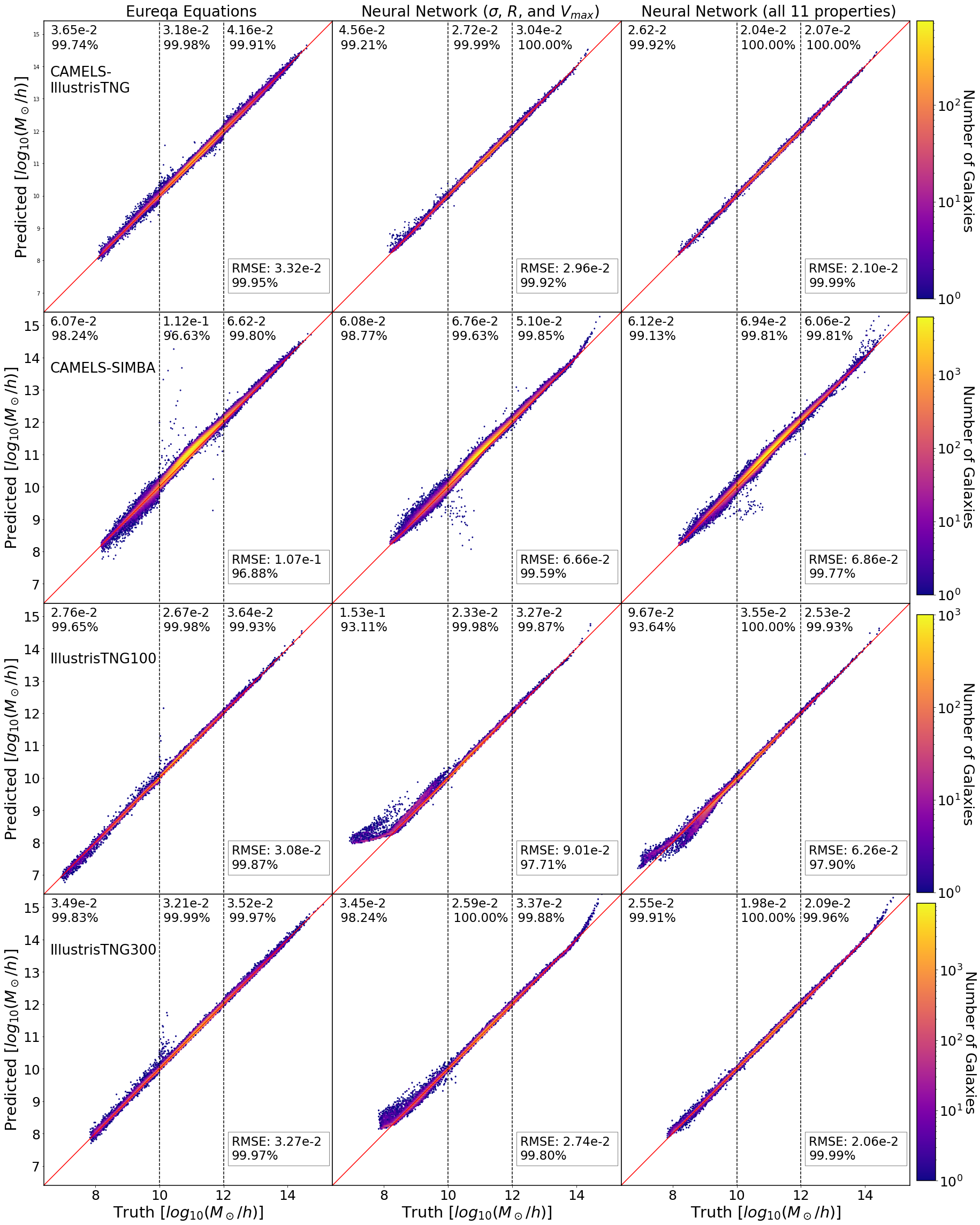}
    \caption{Same as Fig. \ref{fig:pysr_all} but using Eq. \ref{Eq:eureqa} derived from \textsc{Eureqa}.}
    \label{fig:eureqa} 
\end{figure*}

These equations, while more complex, are able to predict the total mass with higher accuracy than the analytic expressions we found using linear regression fitting (Equation \ref{Eq:formula}), with a RMSE value of $3.32 \times 10^{-2}$ for the CAMELS-IllustrisTNG suhalos. When we test this on the CAMELS-SIMBA, IllustrisTNG100, and IllustrisTNG300 simulations we find that the equations are able to effectively extrapolate to the different simulations and to mass ranges different than those used to train the model. The results for this are presented in the panels of the first column of Figure \ref{fig:eureqa}. As shown, the RMSE values for these predictions are lower than those of Equation \ref{Eq:formula} for all four simulations. Similar to Equation \ref{Eq:formula}, we see that the symbolic regression models are able to accurately predict the total mass in the high and low mass regimes without generating biases as the neural networks do. However, Equation \ref{Eq:eureqa} still does not attain the overall low RMSE of the neural networks trained using the same three variables. From this, we conclude that the neural networks are employing a relation that is more complex than the models found by symbolic regression.

\bibliography{references}{}

\begin{thebibliography}{}
\expandafter\ifx\csname natexlab\endcsname\relax\def\natexlab#1{#1}\fi
\providecommand{\url}[1]{\href{#1}{#1}}
\providecommand{\dodoi}[1]{doi:~\href{http://doi.org/#1}{\nolinkurl{#1}}}
\providecommand{\doeprint}[1]{\href{http://ascl.net/#1}{\nolinkurl{http://ascl.net/#1}}}
\providecommand{\doarXiv}[1]{\href{https://arxiv.org/abs/#1}{\nolinkurl{https://arxiv.org/abs/#1}}}

\bibitem[{Akiba {et~al.}(2019)Akiba, Sano, Yanase, Ohta, \& Koyama}]{Optuna}
Akiba, T., Sano, S., Yanase, T., Ohta, T., \& Koyama, M. 2019, in Proceedings
  of the 25rd {ACM} {SIGKDD} International Conference on Knowledge Discovery
  and Data Mining

\bibitem[{{Angl{\'e}s-Alc{\'a}zar}
  {et~al.}(2017{\natexlab{a}}){Angl{\'e}s-Alc{\'a}zar}, {Dav{\'e}},
  {Faucher-Gigu{\`e}re}, {{\"O}zel}, \&
  {Hopkins}}]{Angles-Alcazar2017_BHfeedback}
{Angl{\'e}s-Alc{\'a}zar}, D., {Dav{\'e}}, R., {Faucher-Gigu{\`e}re}, C.-A.,
  {{\"O}zel}, F., \& {Hopkins}, P.~F. 2017{\natexlab{a}}, \mnras, 464, 2840,
  \dodoi{10.1093/mnras/stw2565}

\bibitem[{{Angl{\'e}s-Alc{\'a}zar}
  {et~al.}(2017{\natexlab{b}}){Angl{\'e}s-Alc{\'a}zar}, {Faucher-Gigu{\`e}re},
  {Kere{\v s}}, {Hopkins}, {Quataert}, \&
  {Murray}}]{Angles-Alcazar2017_BaryonCycle}
{Angl{\'e}s-Alc{\'a}zar}, D., {Faucher-Gigu{\`e}re}, C.-A., {Kere{\v s}}, D.,
  {et~al.} 2017{\natexlab{b}}, \mnras, 470, 4698, \dodoi{10.1093/mnras/stx1517}

\bibitem[{{Bender} {et~al.}(1992){Bender}, {Burstein}, \&
  {Faber}}]{galaxy_properties}
{Bender}, R., {Burstein}, D., \& {Faber}, S.~M. 1992, \apj, 399, 462,
  \dodoi{10.1086/171940}

\bibitem[{Burstein {et~al.}(1997)Burstein, Bender, Faber, \&
  Nolthenius}]{stellar_systems}
Burstein, D., Bender, R., Faber, S., \& Nolthenius, R. 1997, The Astronomical
  Journal, 114, 1365, \dodoi{10.1086/118570}

\bibitem[{{Dav{\'e}} {et~al.}(2019){Dav{\'e}}, {Angl{\'e}s-Alc{\'a}zar},
  {Narayanan}, {Li}, {Rafieferantsoa}, \& {Appleby}}]{SIMBA}
{Dav{\'e}}, R., {Angl{\'e}s-Alc{\'a}zar}, D., {Narayanan}, D., {et~al.} 2019,
  \mnras, 486, 2827, \dodoi{10.1093/mnras/stz937}

\bibitem[{{Dressler} {et~al.}(1987){Dressler}, {Lynden-Bell}, {Burstein},
  {Davies}, {Faber}, {Terlevich}, \& {Wegner}}]{dressler}
{Dressler}, A., {Lynden-Bell}, D., {Burstein}, D., {et~al.} 1987, \apj, 313,
  42, \dodoi{10.1086/164947}

\bibitem[{Faber \& Jackson(1976)}]{faberjackson}
Faber, S.~M., \& Jackson, R.~E. 1976, The Astrophysical Journal, 204, 668

\bibitem[{{Hopkins}(2015)}]{Hopkins2015_Gizmo}
{Hopkins}, P.~F. 2015, \mnras, 450, 53, \dodoi{10.1093/mnras/stv195}

\bibitem[{{Loshchilov} \& {Hutter}(2017)}]{AdamW}
{Loshchilov}, I., \& {Hutter}, F. 2017, arXiv e-prints, arXiv:1711.05101.
\newblock \doarXiv{1711.05101}

\bibitem[{{Marinacci} {et~al.}(2017){Marinacci}, {Vogelsberger}, {Pakmor},
  {Torrey}, {Springel}, {Hernquist}, {Nelson}, {Weinberger}, {Pillepich},
  {Naiman}, \& {Genel}}]{MarinacciF_17a}
{Marinacci}, F., {Vogelsberger}, M., {Pakmor}, R., {et~al.} 2017, ArXiv
  e-prints, 1707.03396.
\newblock \doarXiv{1707.03396}

\bibitem[{{Muratov} {et~al.}(2015){Muratov}, {Kere{\v s}},
  {Faucher-Gigu{\`e}re}, {Hopkins}, {Quataert}, \& {Murray}}]{Muratov2015}
{Muratov}, A.~L., {Kere{\v s}}, D., {Faucher-Gigu{\`e}re}, C.-A., {et~al.}
  2015, \mnras, 454, 2691, \dodoi{10.1093/mnras/stv2126}

\bibitem[{{Naiman} {et~al.}(2017){Naiman}, {Pillepich}, {Springel Enrico
  Ramirez-Ruiz}, {Torrey}, {Vogelsberger}, {Pakmor}, {Nelson}, {Marinacci},
  {Hernquist}, {Weinberger}, \& {Genel}}]{NaimanJ_17a}
{Naiman}, J.~P., {Pillepich}, A., {Springel Enrico Ramirez-Ruiz}, V., {et~al.}
  2017, ArXiv e-prints, 1707.03401.
\newblock \doarXiv{1707.03401}

\bibitem[{{Nelson} {et~al.}(2018){Nelson}, {Pillepich}, {Springel},
  {Weinberger}, {Hernquist}, {Pakmor}, {Genel}, {Torrey}, {Vogelsberger},
  {Kauffmann}, {Marinacci}, \& {Naiman}}]{NelsonD_17a}
{Nelson}, D., {Pillepich}, A., {Springel}, V., {et~al.} 2018, \mnras, 475, 624,
  \dodoi{10.1093/mnras/stx3040}

\bibitem[{{Nelson} {et~al.}(2019){Nelson}, {Springel}, {Pillepich},
  {Rodriguez-Gomez}, {Torrey}, {Genel}, {Vogelsberger}, {Pakmor}, {Marinacci},
  {Weinberger}, {Kelley}, {Lovell}, {Diemer}, \&
  {Hernquist}}]{IllustrisTNG_public}
{Nelson}, D., {Springel}, V., {Pillepich}, A., {et~al.} 2019, Computational
  Astrophysics and Cosmology, 6, 2, \dodoi{10.1186/s40668-019-0028-x}

\bibitem[{{Pillepich} {et~al.}(2018{\natexlab{a}}){Pillepich}, {Nelson},
  {Hernquist}, {Springel}, {Pakmor}, {Torrey}, {Weinberger}, {Genel}, {Naiman},
  {Marinacci}, \& {Vogelsberger}}]{Pillepich_2018}
{Pillepich}, A., {Nelson}, D., {Hernquist}, L., {et~al.} 2018{\natexlab{a}},
  \mnras, 475, 648, \dodoi{10.1093/mnras/stx3112}

\bibitem[{{Pillepich} {et~al.}(2018{\natexlab{b}}){Pillepich}, {Springel},
  {Nelson}, {Genel}, {Naiman}, {Pakmor}, {Hernquist}, {Torrey}, {Vogelsberger},
  {Weinberger}, \& {Marinacci}}]{PillepichA_16a}
{Pillepich}, A., {Springel}, V., {Nelson}, D., {et~al.} 2018{\natexlab{b}},
  \mnras, 473, 4077, \dodoi{10.1093/mnras/stx2656}

\bibitem[{{Somerville} \& {Dav{\'e}}(2015)}]{SomervilleDave2015}
{Somerville}, R.~S., \& {Dav{\'e}}, R. 2015, \araa, 53, 51,
  \dodoi{10.1146/annurev-astro-082812-140951}

\bibitem[{{Springel} {et~al.}(2001){Springel}, {White}, {Tormen}, \&
  {Kauffmann}}]{Subfind}
{Springel}, V., {White}, S.~D.~M., {Tormen}, G., \& {Kauffmann}, G. 2001,
  \mnras, 328, 726, \dodoi{10.1046/j.1365-8711.2001.04912.x}

\bibitem[{{Springel} {et~al.}(2018){Springel}, {Pakmor}, {Pillepich},
  {Weinberger}, {Nelson}, {Hernquist}, {Vogelsberger}, {Genel}, {Torrey},
  {Marinacci}, \& {Naiman}}]{SpringelV_17a}
{Springel}, V., {Pakmor}, R., {Pillepich}, A., {et~al.} 2018, \mnras, 475, 676,
  \dodoi{10.1093/mnras/stx3304}

\bibitem[{{Tully} {et~al.}(1975){Tully}, {de Marseille}, \&
  {Fisher}}]{Tully_Fisher}
{Tully}, R.~B., {de Marseille}, O., \& {Fisher}, J.~R. 1975, in Bulletin of the
  American Astronomical Society, Vol.~7, 426

\bibitem[{{Villaescusa-Navarro} {et~al.}(2020){Villaescusa-Navarro}, {Wandelt},
  {Angl{\'e}s-Alc{\'a}zar}, {Genel}, {Zorrilla Mantilla}, {Ho}, \&
  {Spergel}}]{Villaescusa-Navarro_2020c}
{Villaescusa-Navarro}, F., {Wandelt}, B.~D., {Angl{\'e}s-Alc{\'a}zar}, D.,
  {et~al.} 2020, arXiv e-prints, arXiv:2011.05992.
\newblock \doarXiv{2011.05992}

\bibitem[{{Villaescusa-Navarro} {et~al.}(2021){Villaescusa-Navarro},
  {Angl{\'e}s-Alc{\'a}zar}, {Genel}, {Spergel}, {Somerville}, {Dave},
  {Pillepich}, {Hernquist}, {Nelson}, {Torrey}, {Narayanan}, {Li}, {Philcox},
  {La Torre}, {Maria Delgado}, {Ho}, {Hassan}, {Burkhart}, {Wadekar},
  {Battaglia}, {Contardo}, \& {Bryan}}]{CAMELS}
{Villaescusa-Navarro}, F., {Angl{\'e}s-Alc{\'a}zar}, D., {Genel}, S., {et~al.}
  2021, \apj, 915, 71, \dodoi{10.3847/1538-4357/abf7ba}

\bibitem[{{Weinberger} {et~al.}(2019){Weinberger}, {Springel}, \&
  {Pakmor}}]{Arepo_public}
{Weinberger}, R., {Springel}, V., \& {Pakmor}, R. 2019, arXiv e-prints,
  arXiv:1909.04667.
\newblock \doarXiv{1909.04667}

\bibitem[{{Weinberger} {et~al.}(2017){Weinberger}, {Springel}, {Hernquist},
  {Pillepich}, {Marinacci}, {Pakmor}, {Nelson}, {Genel}, {Vogelsberger},
  {Naiman}, \& {Torrey}}]{WeinbergerR_16a}
{Weinberger}, R., {Springel}, V., {Hernquist}, L., {et~al.} 2017, \mnras, 465,
  3291, \dodoi{10.1093/mnras/stw2944}

\end{thebibliography}
\bibliographystyle{aasjournal}

\end{document}